\documentclass[twocolumn,letterpaper,amsmath,amssymb,floatfix,aps,superscriptaddress,nofootinbib]{revtex4}

\usepackage{graphicx}
\usepackage{dcolumn}
\usepackage{bm}
\usepackage[usenames]{color}
\usepackage{hyperref}
\usepackage{ulem}

\begin{document}

\title{Theory of pore-driven and end-pulled polymer translocation dynamics through a nanopore: An overview}

\author{Jalal Sarabadani}
\email{j.sarabadani@lboro.ac.uk}
\affiliation{School of Nano Science, Institute for Research in Fundamental Sciences (IPM), 19395-5531, Tehran, Iran}
\affiliation{Interdisciplinary Centre for Mathematical Modelling, Loughborough University, Loughborough, Leicestershire LE11 3TU, UK}
\affiliation{Department of Mathematical Sciences, Loughborough University, Loughborough, Leicestershire LE11 3TU, UK}
\affiliation{Department of Applied Physics and QTF Center of Excellence, Aalto University School of Science, P.O. Box 11000, FI-00076 Aalto, Espoo, Finland}

\author{Tapio Ala-Nissila}
\affiliation{Department of Applied Physics and QTF Center of Excellence, Aalto University School of Science, P.O. Box 11000, FI-00076 Aalto, Espoo, Finland}
\affiliation{Interdisciplinary Centre for Mathematical Modelling, Loughborough University, Loughborough, Leicestershire LE11 3TU, UK}
\affiliation{Departments of Mathematical Sciences and Physics, Loughborough University,
Loughborough, Leicestershire LE11 3TU, UK}

\begin{abstract}
We review recent progress on the theory of dynamics of polymer translocation through a nanopore based on the iso-flux 
tension propagation (IFTP) theory. We investigate both pore-driven translocation of flexible and a semi-flexible 
polymers, and the end-pulled case of flexible chains
by means of the IFTP theory and extensive molecular dynamics (MD) simulations.
The validity of the IFTP theory can be quantified by the waiting time distributions of the monomers
which reveal the details of the dynamics of the translocation process.  
The IFTP theory allows a parameter-free description of the translocation process and
can be used to derive exact analytic scaling forms in the appropriate limits, 
including the influence due to the pore friction that appears as a finite-size correction to asymptotic scaling. 
We show that in the case of pore-driven semi-flexible and end-pulled polymer chains the IFTP theory must be 
augmented with an explicit {\it trans} side friction term for a quantitative description of the translocation process.
\end{abstract}

\maketitle


\section{Introduction} \label{introduction}

Since the seminal works by Bezrukov {\it et al}. \cite {Parsegian} and two years later by Kasianowicz {\it et al}. \cite{Kasianowicz} in 1996,
translocation of a polymer through a nanopore has become one of the most active research areas in soft matter and biological 
physics \cite{Muthukumar_book,Milchev_JPCM,Tapio_review}.
It has many applications in medicine, biological and soft matter physics and engineering, such as protein transportation through membrane channels
and virus injection \cite{Alberts}.
A setup based on the polymer translocation through a nanopore has been suggested as an inexpensive and rapid method for DNA and other biopolymer sequencing.
Therefore, motivated by these applications many experimental and theoretical works have been focused on the study of the dynamics of the translocation
\cite{Parsegian,Kasianowicz,Muthukumar_book,Milchev_JPCM,Tapio_review,Alberts,dimarzio1979,sung1996,muthu1999,%
chuang2001,Meller_PRL_2001,metzler2003,meller2003,branton_PRL_2003,BatesBioPhysJ2003,kantor2004,meller_biophys_j_2004,%
RehlingFlickeringPore,storm2005,Keyser_NatPhys_2006,grosberg2006,ClementiPRL2006,dubbeldam2007,sakaue2007,Huopaniemi_2007,%
HuhNature2007,branton2008,luo2008,sakaue2008,Aksimentiev_NanoLett_2008,Tapio_PRL_2008,slater2008a,slater2008b,Panja_2008,%
luo2009,bhatta2009,unbiased_Slater_3,Keyser_NatPhys_2009,%
Santtu_EPJE_2009,YameenSmall2009,lathropJACS2010OscillatingForce,yamadaFlickeringPore,schadt2010,Sischka,unbiased_Slater_1,%
bhatta2010,sakaue2010,ChenNatMatt2010,golestanianPRL2011,rowghanian2011,AngeliLabonChip2011,fanzioSciRep2012,StefureacOscillatingForce,%
saito2011,dubbeldam2011,saito2012a,saito2012b,ikonen2012a,ikonen2012b,ikonen2012c,unbiased_Slater_2,%
Golestanian_PRX_2012,golestanianJCP2012,dubbeldam2012,ikonen2013,unbiased_Polson,dubbeldam2013,%
dubbeldam2014,jalal2014,Bulushev_NanoLett_2014,MereutaSciRep2014,Spencer2014,Stein2014,suhonen2014,%
Bulushev_NanoLett_2015,jalal2015,fiasconaroPRE2015,sakaue2016,Bulushev_NanoLett_2016,Menais_SciRep_2016,DaiPolymers2016,%
PlesaNatNano2016,jalal2017SR,jalal2017EPL,Postma,Marconi,Micheletti_PNAS_2017,Slater_2018,Menais_PRE_2018}

The three simplest basic translocation scenarios are the unbiased \cite{unbiased_Slater_3,unbiased_Slater_1,unbiased_Slater_2,unbiased_Polson},
pore-driven and end-pulled setups. While in the pore-driven case the driving force is an electric field (arising from a voltage bias between the two sides 
of the membrane) which acts on the monomer(s) inside the pore, for the end-pulled case the polymer is pulled through a nanopore by either
an atomic force microscope (AFM) \cite{Ritort_2006}, or a magnetic or an optical tweezer 
\cite{Keyser_NatPhys_2006,Keyser_NatPhys_2009,Sischka,Bulushev_NanoLett_2014,Bulushev_NanoLett_2015,Bulushev_NanoLett_2016}.
Among the above scenarios the end-pulled case has been suggested to be a good candidate to slow down and control 
the translocation process which is vital to properly identify the nucleotides in DNA sequencing 
\cite{kantor2004,Huopaniemi_2007,Santtu_EPJE_2009,Slater_2018,jalal2017EPL,jalalDoubleSide_2018}. 

Most of the theoretical research to date has focused on the pore-driven case of a fully flexible chain with a constant radius of the nanopore and under a constant
driving force \cite{sakaue2007,bhatta2009,bhatta2010,saito2012a,saito2012b,ikonen2012a,dubbeldam2012,ikonen2013,jalal2014,jalal2017SR,jalal2017EPL}.
Nevertheless, in many interesting practical cases the translocating chains are not fully flexible -- e.g. double-stranded DNA has a
persistence length of $\ell_{\mathrm p} \approx 500$ {\AA}. Therefore, to unravel the influence of stiffness of the chain on the translocation process
the pore-driven case of a semi-flexible polymer with a finite persistence length has theoretically been recently considered \cite{jalal2017SR}. 

On the other hand, the translocation process has also been studied under a time-dependent external driving force 
\cite{BatesBioPhysJ2003,Aksimentiev_NanoLett_2008,StefureacOscillatingForce,ikonen2012c,golestanianJCP2012,fiasconaroPRE2015}.
As an example, Langevin dynamics simulations have been employed to study polymer translocation under a time-dependent 
alternating driving force which shows that at an optimal frequency of the alternating force, resonant activation occurs
if the polymer-pore interaction is attractive \cite{ikonen2012c}.
For the pore driven case with an oscillating driving force there are some biological applications such as translocation of 
$\alpha$-helical and linear peptides through an $\alpha$-hemolysin nanopore in the presence of an AC field \cite{StefureacOscillatingForce}, 
and using an alternating current signal to monitor the DNA escape from an $\alpha$-hemolysin nanopore \cite{lathropJACS2010OscillatingForce}.
Additionally, use of an alternating electric field at the nanopore has been suggested for the DNA sequencing \cite{Aksimentiev_NanoLett_2008}.

In addition, there are some theoretical and experimental works where the width of the pore changes during the course of the translocation process.
For instance it has been shown that in the nucleocytoplasmic transport in eukaryotes, 
the nuclear pore complex plays an important role \cite{yamadaFlickeringPore}. Moreover, the twin-pore protein translocase (TIM22 complex) in the inner membrane of 
the mitocondria can control the exchange of molecules between mitochondria and the rest of cell \cite{RehlingFlickeringPore}. 
On the numerical side, using molecular dynamics (MD) simulations the active translocation of a polymer through a flickering pore has been considered
and it has been shown that more efficient translocation as compared to the static pore can be obtained when the pore has an alternating 
width and sticky walls \cite{golestanianPRL2011}.
Experiments have confirmed that by applying mechanical stress the cross section of an elastomeric nanochannel device is changed 
and this can modulate the translocation of DNA through the nanochannel \cite{HuhNature2007,AngeliLabonChip2011,fanzioSciRep2012}.
Interestingly, one can tune the width of the nanopore by covering the inside of the nanopore with thermally driven nanoactuation 
of polyNIPAM brushes \cite{YameenSmall2009}.

Over the last few years a quantitative theory for both the pore-driven and end-pulled translocation dynamics of a polymer through a nanopore
has been developed \cite{ikonen2012a,ikonen2012b,ikonen2012c,ikonen2013,jalal2014,jalal2015,jalal2017SR,jalal2017EPL} based on the idea of tension
propagation by Sakaue in 2007 \cite{sakaue2007}. 
The basic picture is that the translocation process constitutes two stages, which are the tension propagation (TP) and post propagation (PP) stages. 
During the TP stage a tension propagates along the backbone of the chain, and the {\it cis} side subchain can be divided into mobile and immobile parts. 
In the mobile part the monomers have already experienced the tension while the rest of the chain is immobile which means that the average velocity of the 
monomers for this part is zero, i.e. that part of the chain is in equilibrium. 
In the PP stage the tension has reached the chain end and 
the whole {\it cis} side subchain is moving towards the nanopore.
Based on the picture above, it has been shown that the time-dependent friction due to the mobile {\it cis} side subchain plays the key role
in translocation dynamics. All these ideas have been amalgamated into a quantitative, parameter-free Iso-Flux Tension Propagation (IFTP) 
theory \cite{jalal2014} by combining tension propagation with the iso-flux assumption \cite{rowghanian2011}.
The IFTP theory describes translocation dynamics solely in terms of the translocation coordinate of the chain, friction due to
the pore and the tension front on the {\it cis} side and it allows exact analytic solutions to the scaling of the
average translocation time as a function of the chain length in some limits.
Recently, the IFTP theory has been augmented to describe the case of semi-flexible polymer chains \cite{jalal2017SR}. 
It has been shown that an additional time-dependent {\it trans} side friction due to stiffness of the chain plays a role in the dynamics 
of the translocation process specially in the short chain limit \cite{jalal2017SR}. 
Moreover, the IFTP theory has been extended to describe end-pulled polymer translocation 
through a nanopore \cite{jalal2017EPL} when the drive is very strong. An exact scaling form for the 
the translocation time as a function of the chain length and the driving force has been derived. 
This extended version of the IFTP theory is in excellent agreement with the MD simulations of coarse-grained polymer chains.

In this paper we present an overview on the current status of the theory of polymer translocation.
We first review the IFTP theory in brief in Sec.~\ref{model} focusing on the limit of strong driving where
analytic results can be derived for the scaling of the average translocation time.
Then we apply the theory to the pore-driven case under a constant driving force and through a static nanopore in Sec.~%
\ref{Sec_pore-driven_flexible_static-pore_constant-force}.
Next, in Sec.~\ref{pore-driven_semi-flexible_static-pore_constant-force} the pore-driven translocation of a semi-flexible chain
is described. After that, the end-pulled case for a flexible chain again under a constant force 
and through a static pore is studied in Sec.~\ref{end-pulled_flexible_static-pore_constant-force}. 
Finally, Sec.~\ref{pore-driven_flexible_flickering-pore_alternating-force} is devoted 
to investigating the pore-driven case under an alternating driving force through a flickering pore.
Summary, conclusions and future prospects are presented in Sec.~\ref{conclusion}.


\section{Theory} \label{model}

For brevity, we use here dimensionless units denoted by tilde 
as $\tilde{Y} \equiv Y / Y_{\rm u}$, with the units of time $t_{\rm u} \equiv \eta a^2 / (k_{\mathrm B} T)$, 
length $s_{\rm u} \equiv a$, velocity $v_{\rm u} \equiv a/t_{\rm u} = k_{\mathrm B} T/(\eta a)$, force $f_{\rm u} \equiv k_{\mathrm B} T/a$, 
friction $\Gamma_{\rm u} \equiv \eta$, and monomer flux $\phi_{\rm u} \equiv k_{\mathrm B} T/(\eta a^2)$,
where $k_{\mathrm B}$ is the Boltzmann constant, $T$ is the temperature of the system, $a$ is the segment 
length, and $\eta$ is the solvent friction per monomer. All parameters without tilde are 
expressed in Lennard-Jones units, which are used in the MD simulations.

\begin{figure*}[t]\begin{center}
    \begin{minipage}[b]{1.0\textwidth}\begin{center}
        \includegraphics[width=1.0\textwidth]{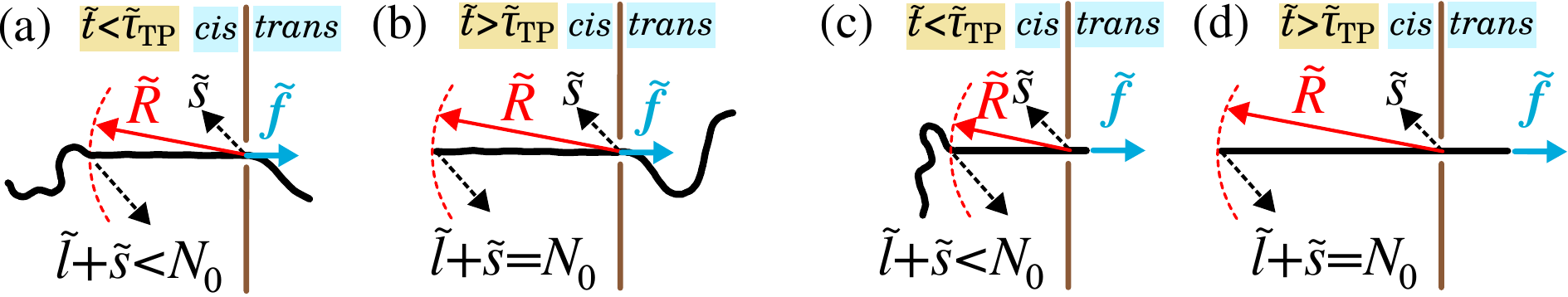}
    \end{center}\end{minipage}
\caption{(a) Schematic of the translocation process during TP stage for the pore-driven case of strong stretching (SS) regime.
$\tilde{f}$ is the driving force and acts only on the monomer(s) inside the pore towards the {\it trans} side. 
$N_0$ and $\tilde{s}$ are the contour length of the polymer and the length of the {\it trans} side subchain, respectively. 
$\tilde{l} + \tilde{s}$ is the number of monomers influenced by the tension force in the {\it cis} side 
which is smaller than the number of total monomers in the polymer $N_0$ during the TP stage. The location 
of the tension front is determined by $\tilde{R}$. (b) The translocation process during the PP stage where 
the tension front has reached the end of the chain, and therefore $\tilde{l} + \tilde{s} = N_0$. 
Panels (c) and (d) are the same as panels (a) and (b), respectively, but for the end-pulled case of SS regime.} 
\label{fig:schimatic}
\end{center}
\end{figure*}

The basic theoretical framework for polymer translocation dynamics
is based on a force balance equation which describes a polymer initially located on the {\it cis} side with one end at the pore.
Starting at time $t=0$ the polymer is subjected to an external force that 
pulls it through to the {\it trans} side during time $t=\tau$, whose average defines the
(average) translocation time. We consider here
the deterministic limit of Brownian dynamics (BD) 
in the overdamped regime, following Refs.~\cite{ikonen2012a,ikonen2012b}. This limit is relevant for the case
where the driving force is strong and noise (thermal fluctuations) can be neglected.
The force-balance equation is written 
for the translocation coordinate $\tilde{s}$ that gives the length of the chain on the {\it trans} side. 
The equation reads
\begin{equation}
\tilde{\Gamma} (\tilde{t}) \frac{d \tilde{s}}{ d \tilde{t}} =
\tilde{F} (\tilde{t}) , 
\label{BD_equation}
\end{equation}
where $\tilde{\Gamma} (\tilde{t})$ is the effective friction, $\tilde{F} (\tilde{t})$ is the force which
acts on the monomers either inside the pore for pore-driven polymers (see Fig.~\ref{fig:schimatic}(a)), 
or only on the head monomer of the chain for end-pulled polymers (see Fig.~\ref{fig:schimatic}(b)). 
The effective friction $\tilde{\Gamma} (\tilde{t})$ depends on the pore friction $\tilde{\eta}_{\textrm{p}} (\tilde{t})$ and the 
drag forces on the {\it cis} and the {\it trans} sides. 

To derive the TP equations, we use arguments similar to Rowghanian {\it et al.}~\cite{rowghanian2011}.
We assume that for both the pore-driven and end-pulled cases the flux of monomers, $\tilde{\phi}\equiv d\tilde{s}/d\tilde{t}$, 
on the mobile domain of the chain and through the pore is constant in space, but evolves in time (iso-flux), which imposes
mass conservation~\cite{rowghanian2011}. Indeed, the monomer flux is defined as $\tilde{\sigma}_0 \tilde{v}_0 \equiv d\tilde{s}/d\tilde{t} $, 
where $\tilde{\sigma}_0$ and $\tilde{v}_0$ are the linear monomer density and the velocity of the monomers at the entrance of the pore, respectively.
The boundary between the mobile and immobile domains, which is called the tension front, is located at distance 
$\tilde{x}=-\tilde{R}(\tilde{t})$ from the pore in the {\it cis} side (see Fig. \ref{fig:schimatic}). 
For the pore-driven case of a flexible chain, inside the mobile domain the external driving force is mediated by the chain backbone
from the pore at $\tilde{x}=0$ all the way to the last mobile monomer $N$ located at the tension front.
In contrast, for the semi-flexible chain the driving force is mediated by the backbone of the chain from the mobile part 
of the chain in the {\it trans} side all the way to the pore at $\tilde{x}=0$, and then to the last mobile monomer $N$ located 
at the tension front.
Finally for the end-pulled case, the driving force is mediated by the chain backbone from the head monomer (head of the polymer 
on which the external force acts) in the {\it trans} side all the way to the pore at $\tilde{x}=0$ and then to the last 
mobile monomer $N$ located at the tension front (see Figs. \ref{fig:schimatic}(c) and \ref{fig:schimatic}(d)). 
Inside the mobile domain, the difference of the tension force between points 
$\tilde{x}'$ and $\tilde{x}' + d \tilde{x}'$ is compensated by the viscous friction force experienced by this part of the polymer
due to its movement. This leads to a local force balance 
relation $d \tilde{f}( \tilde{x}' ) = - \tilde{\phi} (\tilde{t}) d \tilde{x}'$ for the differential element 
$d \tilde{x}'$ that is located between $\tilde{x}'$ and $\tilde{x}' + d\tilde{x}'$.
For the end-pulled case, by integrating the local force balance relation, 
$d \tilde{f}( \tilde{x}' ) = - \tilde{\phi} (\tilde{t}) d \tilde{x}'$, 
over the distance from the head monomer to the pore on the {\it trans} side and then from the pore
to $\tilde{x}$ on the {\it cis} side, the tension force is obtained as 
\begin{equation}
\tilde{f}(\tilde{x},\tilde{t}) = \tilde{f}_0 (\tilde{t}) - \tilde{\phi} (\tilde{t}) \tilde{x},
\label{f_as_func_x}
\end{equation}
where 
\begin{equation}
\tilde{f}_0 (\tilde{t}) \equiv \tilde{F} (\tilde{t}) - \tilde{\eta}_{\mathrm{p}} (\tilde{t}) \tilde{\phi}(\tilde{t}) 
- \tilde{\eta}_{\mathrm{TS}} (\tilde{t}) \tilde{\phi}(\tilde{t}),
\label{f0}
\end{equation}
is the force at the pore entrance on the {\it cis} side.
On the other hand, for the pore-driven case a similar procedure is employed to find the tension force. 
The only difference occurs in the so-called {\it trans} 
side friction $\tilde{\eta}_{\mathrm{TS}} (\tilde{t})$ for the flexible chain, which is absorbed into the pore friction $\tilde{\eta}_{\mathrm{p}}$. 
Similar to the end-pulled case, for the pore-driven case of a semi-flexible chain
the explicit form of $\tilde{\eta}_{\mathrm{TS}} (\tilde{t})$ must also be taken 
into account too, as will be explained in subsection \ref{friction_trans_side_semi_flexible}.

Integration of the force balance equation over the mobile domain gives an expression for the monomer flux as a 
function of the force and the linear size of the mobile domain as \cite{jalal2014}
\begin{equation}
\tilde{\phi} (\tilde{t}) = \frac{\tilde{F} (\tilde{t})}
{ \tilde{R} (\tilde{t}) + \tilde{\eta}_{\textrm{p}} (\tilde{t}) + \tilde{\eta}_{\mathrm{TS}} (\tilde{t}) }.
\label{phi_equation}
\end{equation}
Eq. (\ref{BD_equation}), which is an equation of motion for the translocation coordinate $\tilde{s}$ 
that gives its time evolution and the definition of the flux, $\tilde{\phi}\equiv d\tilde{s}/d\tilde{t}$, can be 
then used to find the expression for the effective friction as
\begin{equation}
\tilde{\Gamma} (\tilde{t}) = \tilde{R}(\tilde{t}) + \tilde{\eta}_{\mathrm{p}} (\tilde{t}) + \tilde{\eta}_{\mathrm{TS}} (\tilde{t}) , 
\label{Gamma_equation}
\end{equation}
which nicely reveals the role of the different the friction terms $\tilde{R}(\tilde{t})$, $\tilde{\eta}_{\mathrm{TS}} (\tilde{t})$ and 
$\tilde{\eta}_{\mathrm{p}} (\tilde{t})$ due to the mobile subchain on the {\it cis} and {\it trans} sides, and due to the pore, 
respectively.

Equations (\ref{BD_equation}), (\ref{phi_equation}) and (\ref{Gamma_equation}) determine the time evolution of $\tilde{s}$, 
but the full solution still requires the knowledge of $\tilde{R}(\tilde{t})$. The derivation of the equation of motion 
for $\tilde{R}(\tilde{t})$ must be done separately for the TP and PP stages. 
In the TP stage, the tension has not reached the final monomer as depicted in Fig.~\ref{fig:schimatic}(a). Here the propagation of the tension 
front into the immobile domain is determined by the geometric shape of the immobile domain. To this end the
scaling relation of the end-to-end distance of the chain is needed in order to obtain a closure relation. For flexible self-avoiding chains
this is simply given by
$\tilde{R} = A_{\nu} N^{\nu}$, where $\nu$ is the Flory exponent, $A_\nu$ is a constant prefactor and $N$ is the last monomer inside the tension front.
For semi-flexible chains the scaling form is more complicated and will be discussed in Sec.  \ref{end_to_end_distance_semi_flexible_chain}.
One can then derive an equation of motion for the tension front as
\begin{equation}
\dot{\tilde{R}} (\tilde{t}) = \mathcal{X}_{\mathrm{TP}},
\label{evolution_of_R_propagation}
\end{equation}
where $\mathcal{X}_{\mathrm{TP}}$ can be a function of $\tilde{R}$, $\tilde{\phi}$ and other parameters of the system.
In the PP stage which is illustrated in Figs.~\ref{fig:schimatic}(b) and (d), every monomer on the {\it cis} side is affected 
by the tension force. 
Therefore, we have the condition $N=N_0$. Since $N$ is also equal to the number of monomers already translocated, $\tilde{s}$, 
plus the number of currently mobile monomers on the {\it cis} side, $\tilde{l}$, the correct closure relation for the PP stage 
is $\tilde{l}+\tilde{s}=N_0$. The equation of motion for the tension front is then derived as
\begin{equation}
\dot{\tilde{R}} (\tilde{t}) = \mathcal{X}_{\mathrm{PP}},
\label{evolution_of_R_post_propagation_2}
\end{equation}
where $\mathcal{X}_{\mathrm{PP}}$ can again be function of $\tilde{R}$, $\tilde{\phi}$ and other parameters of the system.

The self-consistent solution for the model in the TP stage can be obtained from Eqs. (\ref{BD_equation}), 
(\ref{phi_equation}), (\ref{Gamma_equation}) and (\ref{evolution_of_R_propagation}), while for the PP stage one uses 
the set of Eqs. (\ref{BD_equation}), (\ref{phi_equation}), (\ref{Gamma_equation}) 
and (\ref{evolution_of_R_post_propagation_2}).


\section{Pore-driven flexible chain} \label{Sec_pore-driven_flexible_static-pore_constant-force}

In this section we briefly review the pore-driven translocation of a flexible chain through a nanopore, 
where the constant external driving force acts solely to the monomer(s) inside the pore. Then the force balance Eq. (\ref{BD_equation}) 
is cast into
\begin{equation}
\tilde{\Gamma} (\tilde{t}) \frac{d \tilde{s}}{ d \tilde{t}} = \tilde{f} ,
\label{BD_eq_felxible-static}
\end{equation}
where $\tilde{f}$ is the constant driving force at the pore.
Here we investigate a static pore which means the radius of the pore is constant. Therefore, in the theory the pore friction is constant, 
i.e. $\tilde{\eta}_{\mathrm{p}} (\tilde{t}) = \tilde{\eta}_{\mathrm{p}}$.
Moreover, the dynamical {\it trans} side contribution to the friction can be absorbed into the constant pore friction $\tilde{\eta}_{\mathrm{p}}$, 
as we have shown in Refs.~\cite{ikonen2012a,ikonen2012b,ikonen2013}. Thus the force at the entrance of 
the pore in the {\it cis} side in Eq. (\ref{f0}) is  
$\tilde{f}_0 (\tilde{t}) \equiv \tilde{f} - \tilde{\eta}_{\mathrm{p}}  \tilde{\phi}(\tilde{t})$, 
and consequently the flux of monomers $\tilde{\phi} (\tilde{t})$ and the effective friction $\tilde{\Gamma} (\tilde{t})$ are 
\begin{equation}
\tilde{\phi} (\tilde{t}) = \frac{ \tilde{f} } { \tilde{R} (\tilde{t}) + \tilde{\eta}_{\textrm{p}} }; 
\hspace{+1.0cm} 
\tilde{\Gamma} (\tilde{t}) = \tilde{R}(\tilde{t}) + \tilde{\eta}_{\mathrm{p}} . 
\label{Phi_Gamma_flexible_static}
\end{equation}

\begin{figure*}[t]\begin{center}
    \begin{minipage}[b]{0.332\textwidth}\begin{center}
        \includegraphics[width=1.0\textwidth]{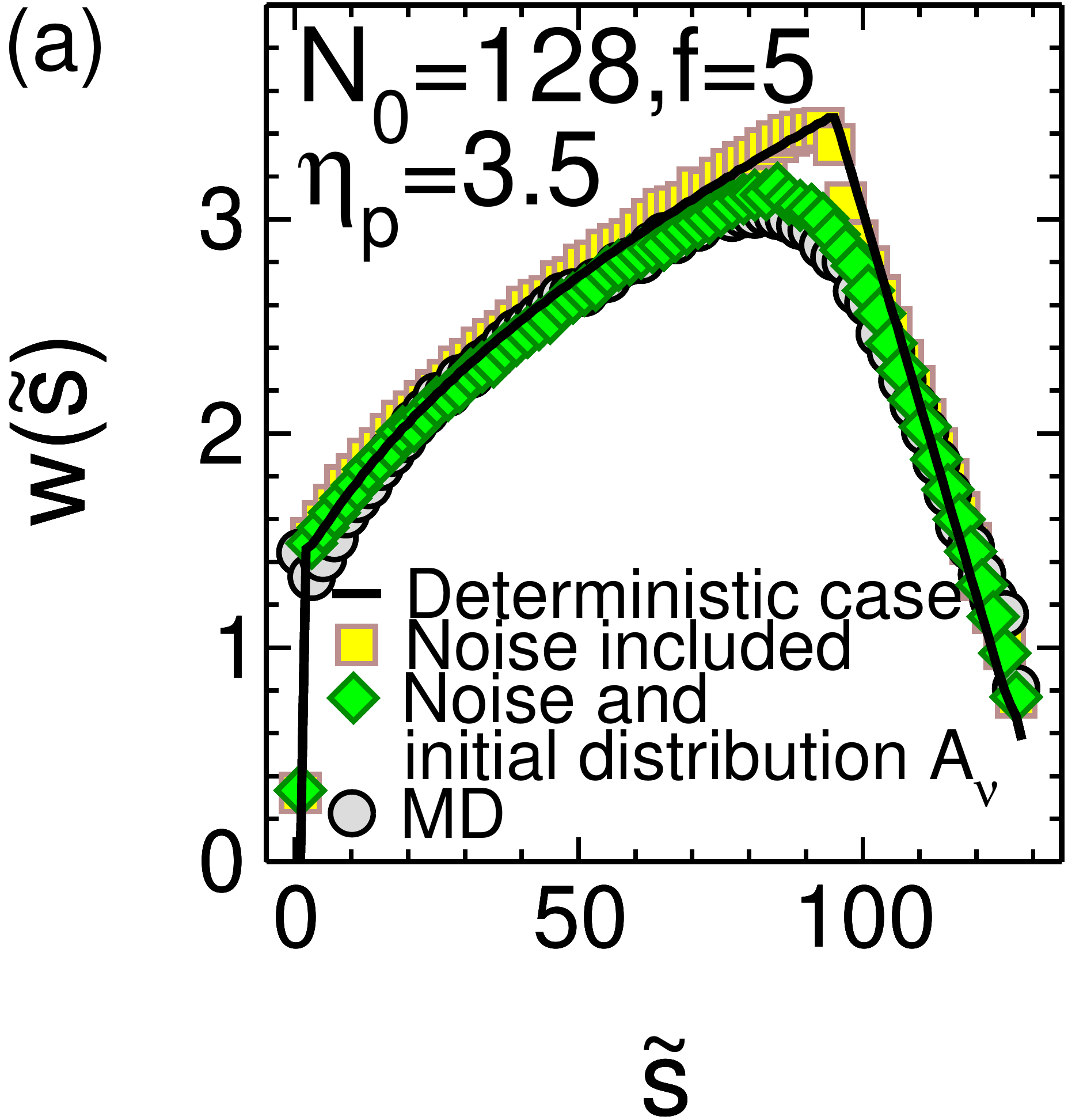}
    \end{center}\end{minipage} \hskip-0.1cm
    \begin{minipage}[b]{0.332\textwidth}\begin{center}
        \includegraphics[width=1.0\textwidth]{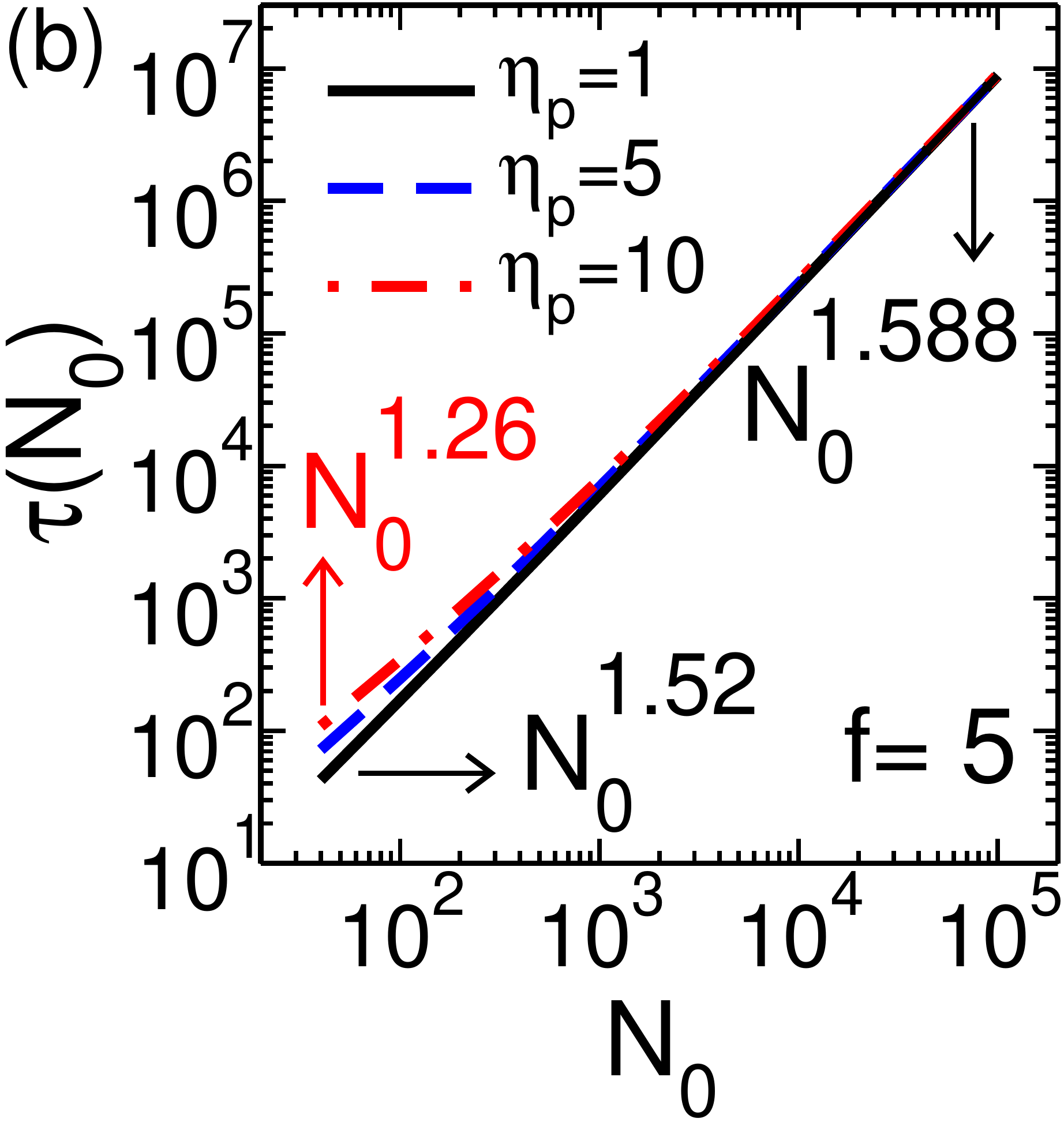}
    \end{center}\end{minipage} \hskip-0.1cm
    \begin{minipage}[b]{0.332\textwidth}\begin{center}
        \includegraphics[width=1.0\textwidth]{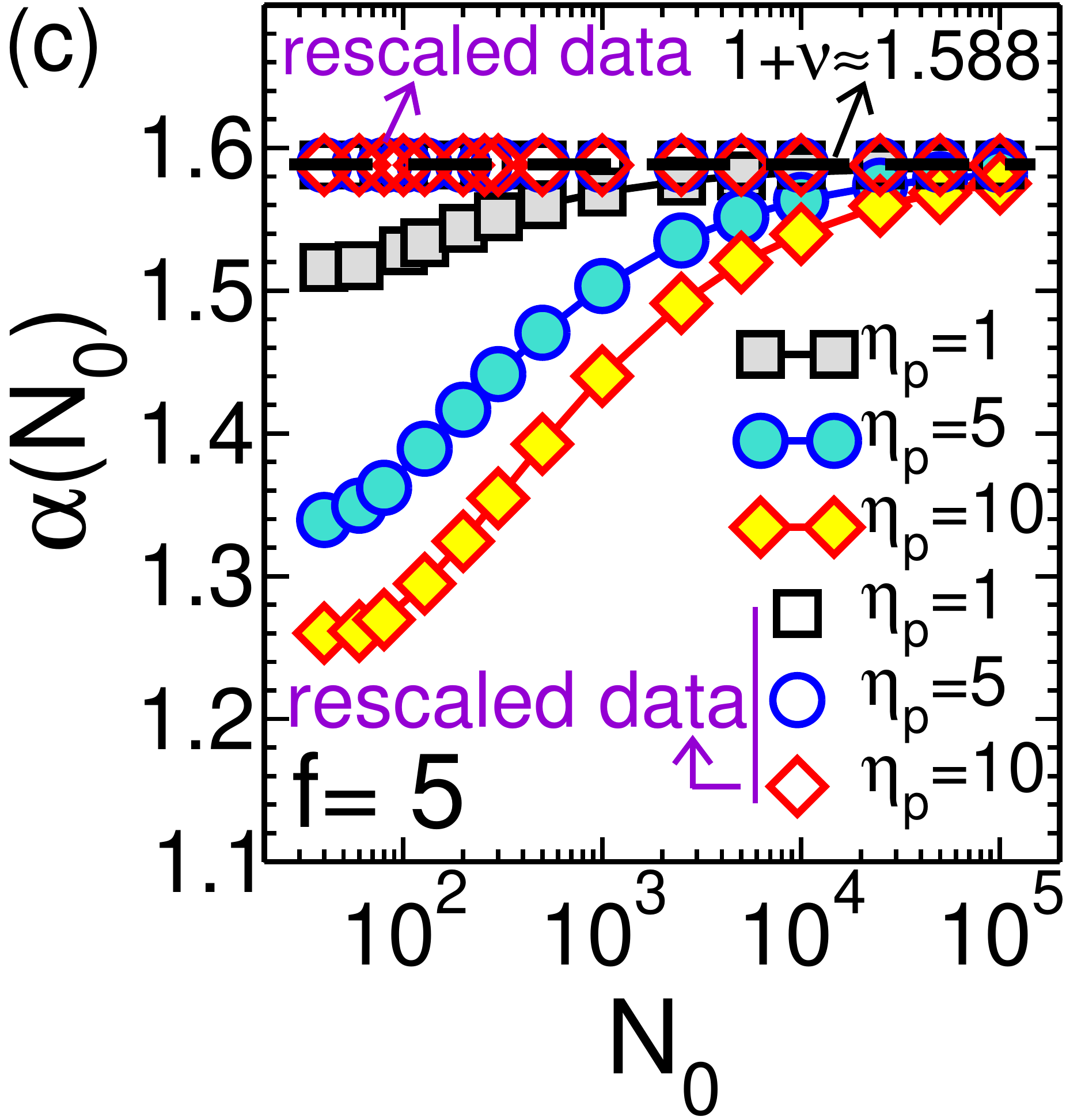}
    \end{center}\end{minipage} \hskip-0.0cm
\caption{(a) The waiting time (WT), $w (\tilde{s})$, as a function of $\tilde{s}$, the translocation coordinate, 
for the pore-driven case of a flexible chain. 
Here, the WT is presented for different cases. The black curve is the WT when both the driving force 
and $A_{\nu}= 1.15$ are deterministic (i.e. fluctuations due to temperature are not taken into account). 
The yellow squares show WT when $A_{\nu}= 1.15$ is deterministic but a stochastic noise term is added in the
driving force in Eq. (8) (see Ref.~\cite{jalal2014} for details).
The green diamonds present the WT when the force is fluctuating and the amplitude $A_{\nu}$ is sampled from 
a distribution generated by thermal fluctuations of the chain with the first bead attached to the entrance of the pore.
Finally, the gray circles are MD simulation data.
(b) The translocation time as a function of the polymer contour length, $N_0$, for fixed values of 
the driving force $f= 5$ and $A_{\nu}= 1.15$, for different values of the pore friction $\eta_{\textrm{p}}= 1, 5$ and 10. 
The effective scaling exponent $1.52$ is for $N_0 = 40$, which is the shortest chain, and smallest chosen pore friction 
$\eta_{\textrm{p}}= 1$, while its value for $\eta_{\textrm{p}}= 10$, which is the highest pore friction, is $1.26$. 
The effective translocation exponent for the longest chain, $N_0 = 1 \times 10^{5}$, is
$\nu + 1 \approx 1.588$. (c) The effective exponent, $\alpha (N_0)$, and the rescaled exponent, as a function of $N_0$ for 
different values of pore friction $\eta_{\textrm{p}} = 1, 5$ and 10.  The rescaled exponents for different 
values of pore friction collapse on a single curve, which is $\alpha^{\dag} (N_0) = 1 + \nu$. See text for details. 
} 
\label{WT_trans_time_pore_driven_flexible_static_fig}
\end{center}
\end{figure*}

As discussed briefly in Sec. \ref{model} 
Eqs. (\ref{BD_eq_felxible-static}) and (\ref{Phi_Gamma_flexible_static}) give the time evolution of the translocation coordinate $\tilde{s}$, 
but to have a full solution the time evolution of the tension front is needed both in the TP and in PP stages.
To this end we only show how the time evolution of the tension front is obtained for the strong stetching (SS)
regime. The same procedure has been applied to the 
trumpet (TR) and stem-flower (SF) regimes, where the external driving force is moderate,
and can be found in Refs. \cite{jalal2014} and \cite{jalal2015}. Here we consider a flexible chain where the distance between 
the tension front and the pore is written as $\tilde{R}_N = A_{\nu} N^{\nu}$, and $N=\tilde{l}+\tilde{s}<N_0$ for the TP stage, and in the SS 
regime $\tilde{l} = \tilde{R} (\tilde{t})$. 
After substituting $\tilde{R} (\tilde{t})+\tilde{s}$ instead of $N$ in the right hand side of the relation above for 
$\tilde{R}_N$ [$\tilde{R} (\tilde{t}) = \tilde{R}_N $], 
and taking the time derivative on both sides, the equation for the time evolution of $\tilde{R} (\tilde{t})$ in the TP stage is written as
\begin{equation}
\dot{\tilde{R}} (\tilde{t}) = \frac{\nu A_{\nu}^{ \frac{1}{\nu} }  \tilde{R} (\tilde{t})^{ \frac{\nu -1}{\nu} }  \tilde{\phi} (\tilde{t}) }
{ 1 -  \nu A_{\nu}^{ \frac{1}{\nu} }  \tilde{R} (\tilde{t})^{ \frac{\nu -1}{\nu} } }.
\label{evolution_of_R_TP_SS_static}
\end{equation}
In the PP stage the tension has reached the chain end and the correct closure relation is $N=\tilde{l}+\tilde{s}=N_0$.
For the SS regime in the PP stage $\tilde{l} = \tilde{R} (\tilde{t})$. By substituting $\tilde{R} (\tilde{t})$ instead of $\tilde{l}$,
the closure relation is $\tilde{R} (\tilde{t}) +\tilde{s}=N_0$. The time derivative on both sides of this new closure relation gives
the time evolution of the tension front in the PP stage and SS regime as
\begin{equation}
\dot{\tilde{R}} (\tilde{t}) = - \tilde{\phi} (\tilde{t}).
\label{evolution_of_R_PP_SS_static}
\end{equation}
Therefore, a self-consistent solution for the model in the TP stage can be obtained from Eqs. (\ref{BD_eq_felxible-static}), 
(\ref{Phi_Gamma_flexible_static}) and (\ref{evolution_of_R_TP_SS_static}), while in the PP stage one uses the set of 
Eqs. (\ref{BD_eq_felxible-static}), (\ref{Phi_Gamma_flexible_static}) and (\ref{evolution_of_R_PP_SS_static}).


\subsection{Waiting time} \label{waiting_time_felxible_static}

To examine the dynamics of the translocation process we focus on one of the most important quantities, the waiting time (WT) distribution, 
which is the time that each segment or monomer spends at the pore during the course of the translocation process. To this end we compare the WT from 
MD with the one from the IFTP theory. For each individual monomer the WT is averaged over many different simulation trajectories.
The details of the MD simulations can be found in Refs. \cite{ikonen2012a} and \cite{jalal2017SR}.

In Fig. \ref{WT_trans_time_pore_driven_flexible_static_fig}(a) we present the WT as a function of the translocation coordinate
$\tilde{s}$ for a fixed chain length $N_0 = 128$, external driving force $f = 5$ and pore friction $\eta_{\textrm{p}} = 3.5$. 
The two stages of the translocation process are clearly revealed in the WT distribution. The first TP stage is where 
more and more mobile monomers are involved in the friction. Consequently the dynamics of the system gets slower and 
therefore the WT grows. The WT gets its maximum when the tension reaches the chain end. In the second PP stage, the tension has already reached the chain end. 
Therefore, all monomers 
of the remaining part of the subchain in the {\it cis} side are mobile and contribute to the friction. When the 
time passes in the PP stage and the {\it cis} subchain is sucked into the pore, the number of mobile monomers 
in the {\it cis} side decreases and consequently the friction decreases, too. Thus the chain speed increases and the WT decreases. 
When the whole chain traverses the nanopore the process of translocation ends.

The IFTP theory can be used to separately examine (a) the influence of the distribution of the initial 
configuration of the chain and (b) thermal fluctuations in the noise. (a) To this end, for chain length $N_0=321$ 
an analytical function for the end-to-end distance distribution of the chain was fitted to MD simulation data. The 
fitting function used was
\begin{equation}
P(z)  = {\mathcal{A}} ~\! z^{\mathcal{B}} {\textrm{exp}} \big[ {\mathcal{C}} z^{\mathcal{D}} ],
\label{distribution_initial_configuration}
\end{equation}
where ${\mathcal{A}}=0.4252$, ${\mathcal{B}}=1.0310$, ${\mathcal{C}}=-1.4417$, ${\mathcal{D}}=2.6203$, and $z$ 
is the normalized end-to-end distance, i.e. $z= \tilde{R} / \langle \tilde{R} \rangle $. It was shown that the 
the same function can be used for shorter chains as well \cite{jalal2014}. 
Many different initial configurations can be sampled by using Eq. (\ref{distribution_initial_configuration}).
The end-to-end distance $\tilde{R}$ is redefined as $\tilde{R} = A_{\nu} (z) ~ N_0^{\nu} $, where $z$ is  
chosen from the probability distribution function in Eq. (\ref{distribution_initial_configuration}).
Then, an approximate distribution of $\tilde{R}$ is incorporated into the IFTP theory through $A_{\nu} (z) = z A_{\nu}$.
For the case (b) to include thermal fluctuations of the noise in the IFTP model, a stochastic force term is added 
to the right hand side of the Eq. (\ref{BD_eq_felxible-static}). A stochastic differential equation for the force balance
is then numerically solved.

In Fig. \ref{WT_trans_time_pore_driven_flexible_static_fig}(a) the WT distribution, $w (\tilde{s})$, is plotted as a function 
of $\tilde{s}$, the translocation coordinate. 
Here, the WT is presented for different levels of stochasticity. The black curve is the WT when both of the driving 
force and $A_{\nu}= 1.15$ are deterministic. The yellow squares show WT when $A_{\nu}= 1.15$ is deterministic but the driving 
force is stochastic, i.e. when the force balance equation includes the thermal noise term. The green diamonds present 
the WT when both the force and $A_{\nu}$ are stochastic. Finally, the gray circles are MD simulation data.

As can be seen in Fig. \ref{WT_trans_time_pore_driven_flexible_static_fig}(a) the transition from the TP to PP stages 
is smoothened by the stochastic sampling of the initial configurations. This feature is also seen in the MD simulations (gray circles),
where we sample the initial configurations by equilibrating the polymer before each actual translocation event.
All in all, there is a very good quantitative agreement between the result of the full stochastic IFTP theory and the MD simulations.


\subsection{Scaling of the translocation time} \label{translocation_time_felxible_static_subsection}

The average translocation time $\tilde{\tau}$ is the most fundamental quantity related to the translocation process.
Here our aim is to present how an analytical form for the translocation time is obtained in the SS regime. The same procedure 
can be applied for the TR and SF regimes \cite{jalal2014,jalal2015} and the final result for the scaling of the translocation time 
for all the SS, TR and SF regimes is the same.

Combining the definition of the flux $\tilde{\phi}= {d\tilde{s}}/{d\tilde{t}}$ and the equation for the flux 
$\tilde{\phi}(\tilde{t})= \tilde{f}/\big[ \tilde{R}(\tilde{t}) + \tilde{\eta}_{\mathrm p} \big]$, together with the mass conservation in the TP stage
$N = \tilde{l} + \tilde{s}$, and $\tilde{l} = \tilde{R}$ for the SS regime, by integration of $N$ from $0$ to $N_0$ the time for the TP stage reads
\begin{equation}
\tilde{\tau}_\mathrm{TP} = \frac{1}{\tilde{f}} \bigg[ \int_0^{N_0} \tilde{R}_N dN + \tilde{\eta}_{\mathrm{p}} N_0 \bigg]
- \Delta \tilde{\tau}_\mathrm{SS},
\label{TP_time_1}
\end{equation}
where $\Delta \tilde{\tau}_\mathrm{SS}=   ( \tilde{R}_{N_0}^2 /2
+ \tilde{\eta}_{\mathrm{p}} \tilde{R}_{N_0})/{\tilde f}$.
In the PP stage the tension has already reached the chain end $N=N_0$ and therefore $dN/d\tilde{t} = 0$.
By integrating $\tilde{R}$ from $\tilde{R}_{N_0}$ to $0$, PP time $\tilde{\tau}_\mathrm{PP}$ is obtained as
\begin{equation}
\tilde{\tau}_\mathrm{PP} = \Delta \tilde{\tau}_\mathrm{SS}.
\end{equation}
Finally, the whole translocation time, $\tilde{\tau} = \tilde{\tau}_\mathrm{TP} + \tilde{\tau}_\mathrm{PP}$, is given by 
\footnote{It should be noted that due to a technical mistake the authors in Ref. \cite{dubbeldam2014} concluded that there is 
term $ \propto N^{2 \nu}$ in the scaling of the translocation time.}
\begin{equation}
\tilde{\tau} = \frac{1}{\tilde{f}} \bigg[ \int_0^{N_0} \hspace{-0.2cm} \tilde{R}_N dN + \tilde{\eta}_{\mathrm{p}} N_0 \bigg]
= \frac{A_\nu N_0^{1+\nu}}{(1+\nu)\tilde{f}} + \frac{\tilde{\eta}_{\mathrm{p}} N_0 }{\tilde{f}} .
\label{translocation_time_felxible_static}
\end{equation}
This analytical result for the translocation time is in excellent agreement with MD simulations and the previous 
scaling analysis in Ref. \cite{ikonen2013}.

According to Eq. (\ref{translocation_time_felxible_static}) and the conventional scaling form, $\tilde{\tau}\propto N_0^\alpha$, 
the effective exponent $\alpha$ is a function of chain length and pore friction. In the language of critical phenomena the second term
on the r.h.s of Eq. (15) could be called a correction-to-scaling term.
To elucidate the influence of the pore friction dependent term the theory has been solved numerically and in 
Fig. \ref{WT_trans_time_pore_driven_flexible_static_fig}(b)
the translocation time, $\tau (N_0)$, has been plotted as a function of the chain length, $N_0$, for fixed values 
$f=5$, $k_{\mathrm B}T=1.2$, $\eta=0.7$, and different values of the pore friction $\eta_{\textrm{p}}= 1, 5$ and 10.
Here we have solved the model deterministically without any stochasticity and used a fixed value of
$A_\nu=1.15$. For short chains the slope depends on the pore friction, while for the long 
chain limit this dependence vanishes and the asymptotic limit is reached where the exponent is $\alpha = 1 + \nu$.
To present the dependence of the effective translocation exponent even more clearly in Fig. \ref{WT_trans_time_pore_driven_flexible_static_fig}(c) 
we have plotted a running translocation exponent, defined as $\alpha (N_0) = \textrm{d} \ln \tau/( \textrm{d} \ln N_0 )$ \cite{ikonen2013}, 
as a function of the chain length for various values of the pore friction $\eta_{\textrm{p}}= 1, 5$ and 10.

The correction-to-scaling term can be actually canceled out by defining a rescaled translocation time as
\begin{equation}
\tilde{\tau}^{\dag} = \tilde{\tau} - a_2 \tilde{\eta}_{\textrm{p}} N_0 = a_1 N_0^{1+\nu} \sim N_0^{\alpha^{\dag}} ,
\label{trans_time_rescaled_as_chain_length_eq}
\end{equation}
where $\alpha^{\dag} \equiv 1 + \nu$ is the rescaled translocation exponent which does not depend on the
chain length any more. We show the rescaled data for different values of the pore friction in 
Fig. \ref{WT_trans_time_pore_driven_flexible_static_fig}(c). The intercept and slope of the curve 
$\tau / N_0^{1+\nu}$ as a function of $\tilde{\eta}_{\textrm{p}} N_0^{-\nu}$ are $a_1$ and $a_2$, respectively, 
as explained in Ref. \cite{ikonen2013} the coefficients $a_1$ and $a_2$ come from a simple linear least squares fit.


\section{Pore-driven semi-flexible chain} \label{pore-driven_semi-flexible_static-pore_constant-force}

So far we have discussed the case where the polymer chain is fully flexible meaning that it follows the simple Flory
scaling form in equilibrium.
However, the DNA is a semi-flexible polymer and we need to reconsider the theory for such chains.
To this end, two crucial points must be investigated. The first one is the possible role of the 
{\it trans} side friction while the second is
the scaling of the end-to-end distance of the semi-flexible chain, which is nontrivial and should comprise the limiting cases of
a rod, an ideal Gaussian chain, as well as an excluded volume chain \cite{Nakanishi}. 

\begin{figure*}[t]\begin{center}
    \begin{minipage}[b]{0.4859\textwidth}\begin{center}
        \includegraphics[width=1.0\textwidth]{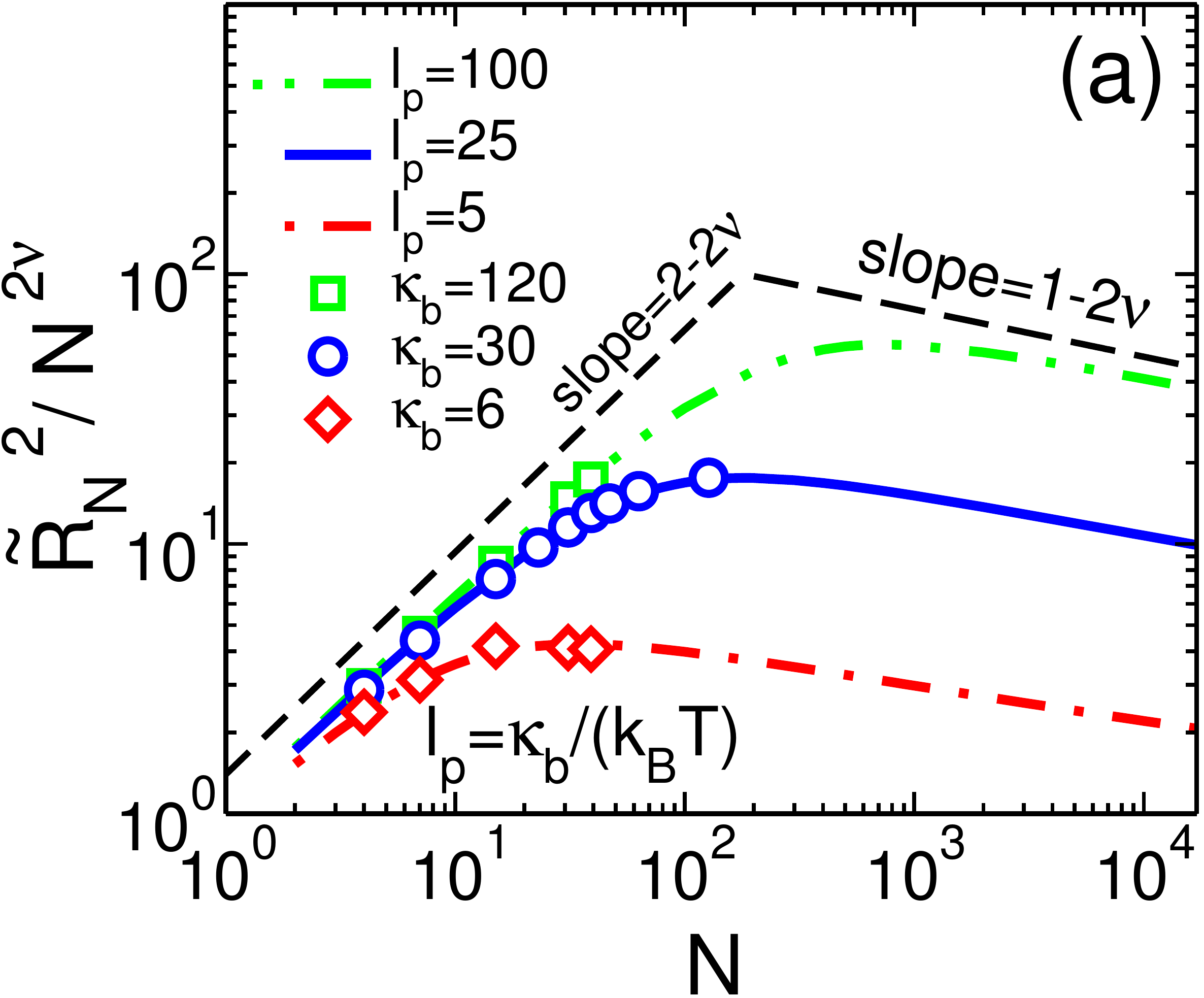}
    \end{center}\end{minipage} \hskip-0.135cm
    \begin{minipage}[b]{0.40\textwidth}\begin{center}
        \includegraphics[width=1.0\textwidth]{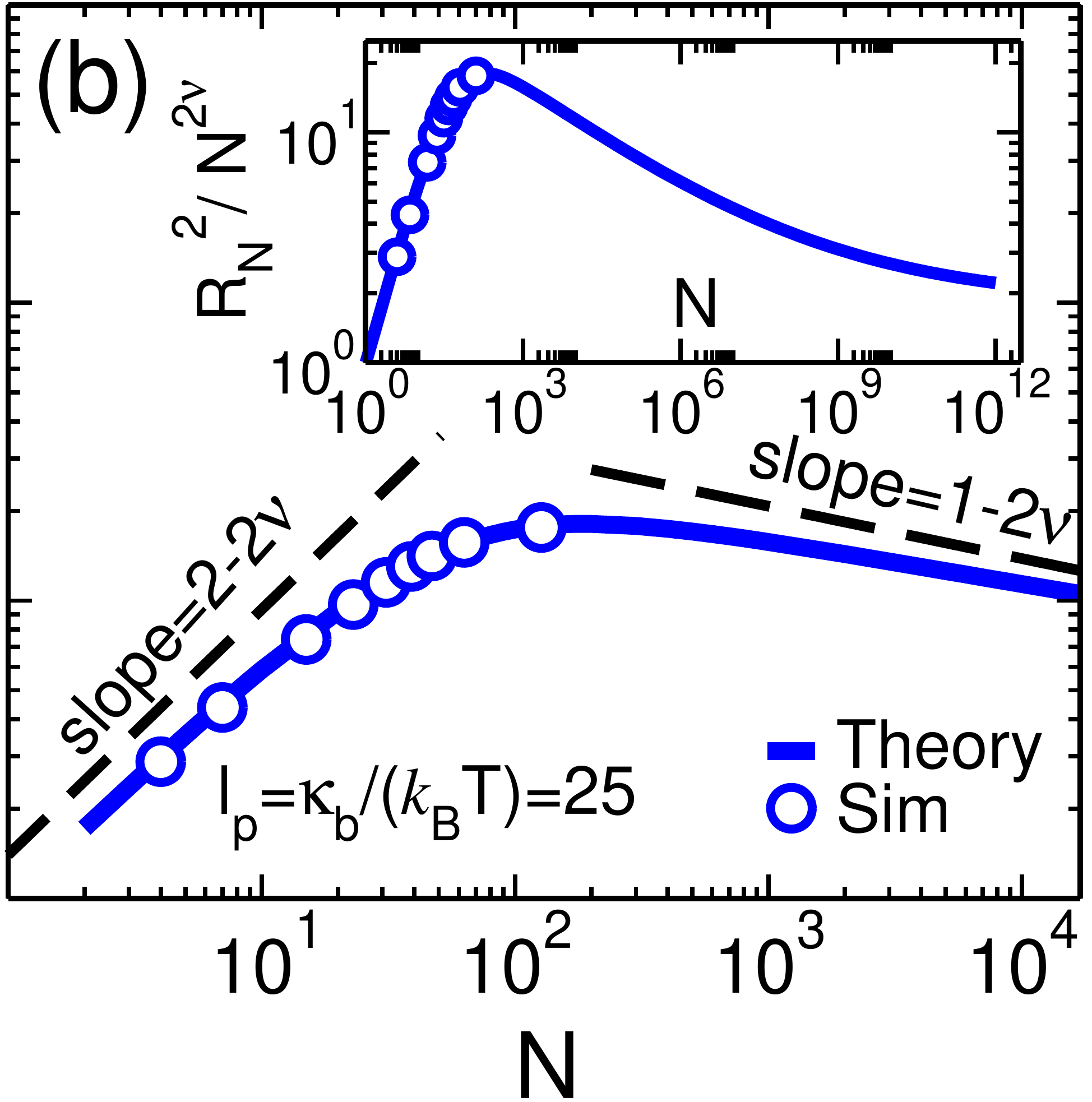}
    \end{center}\end{minipage}
\caption{(a) Normalized end-to-end distance of a semi-flexible chain $\tilde{R}_N^2/N^{2\nu}$ as a 
function of the chain length $N$ for fixed $k_{\mathrm B} T=1.2$ and different values of the bending rigidity (in the MD simulations) 
$\kappa_{{\mathrm b}}= 6$ (red diamonds), 30 (blue circles) and 120 (green squares), which correspond 
to $\ell_{\mathrm p}= 5$ (red dashed-dotted line), 25 (blue solid line) and 100 (green dashed-dotted-dotted line), respectively, according to 
$\ell_{{\mathrm p}}= \kappa_{{\mathrm b}}/(k_{\mathrm B} T)$ in 3D.
The lines come from the analytical interpolation formula of Eq. (\ref{end_to_end_distance}). 
(b) Main panel is the same as (a) but only for bending rigidity 30 and $\ell_{\mathrm p}= 25$. 
To present the asymptotic behavior of the end-to-end distance, $\tilde{R}_N/N^{2\nu}$ has been plotted for an extended range of $N$ in the inset.
It shows that eventually it crosses from a Gaussian intermediate range to a self-avoiding chain at very large $N/\tilde{\ell}_{\mathrm p}$.
} 
\label{end-to-end_pore_driven_flexible_static_fig}
\end{center}
\end{figure*}
\begin{figure*}[t]\begin{center}
    \begin{minipage}[b]{0.426\textwidth}\begin{center}
        \includegraphics[width=1.0\textwidth]{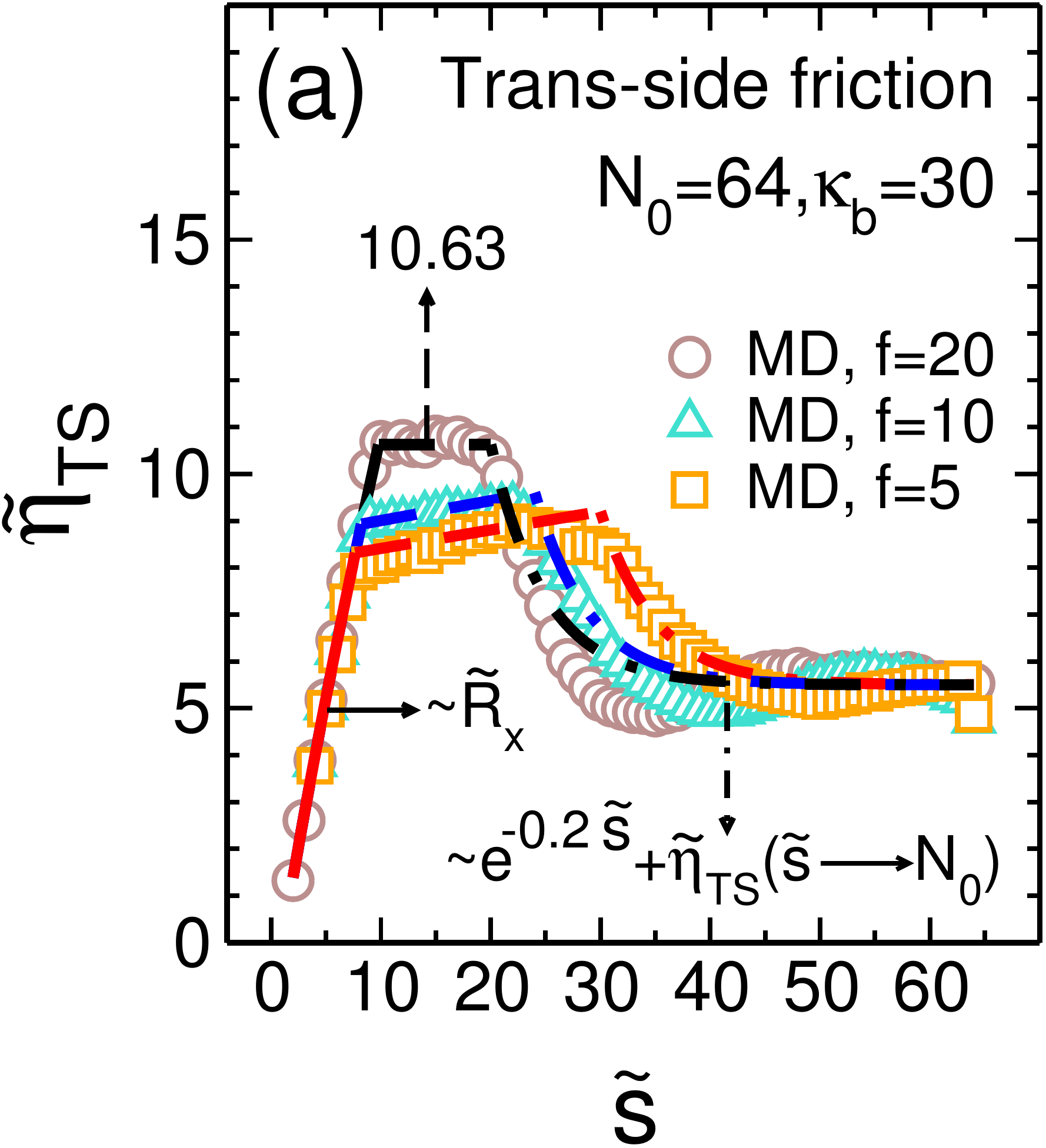}
    \end{center}\end{minipage} \hskip-0.14cm
    \begin{minipage}[b]{0.336\textwidth}\begin{center}
        \includegraphics[width=1.0\textwidth]{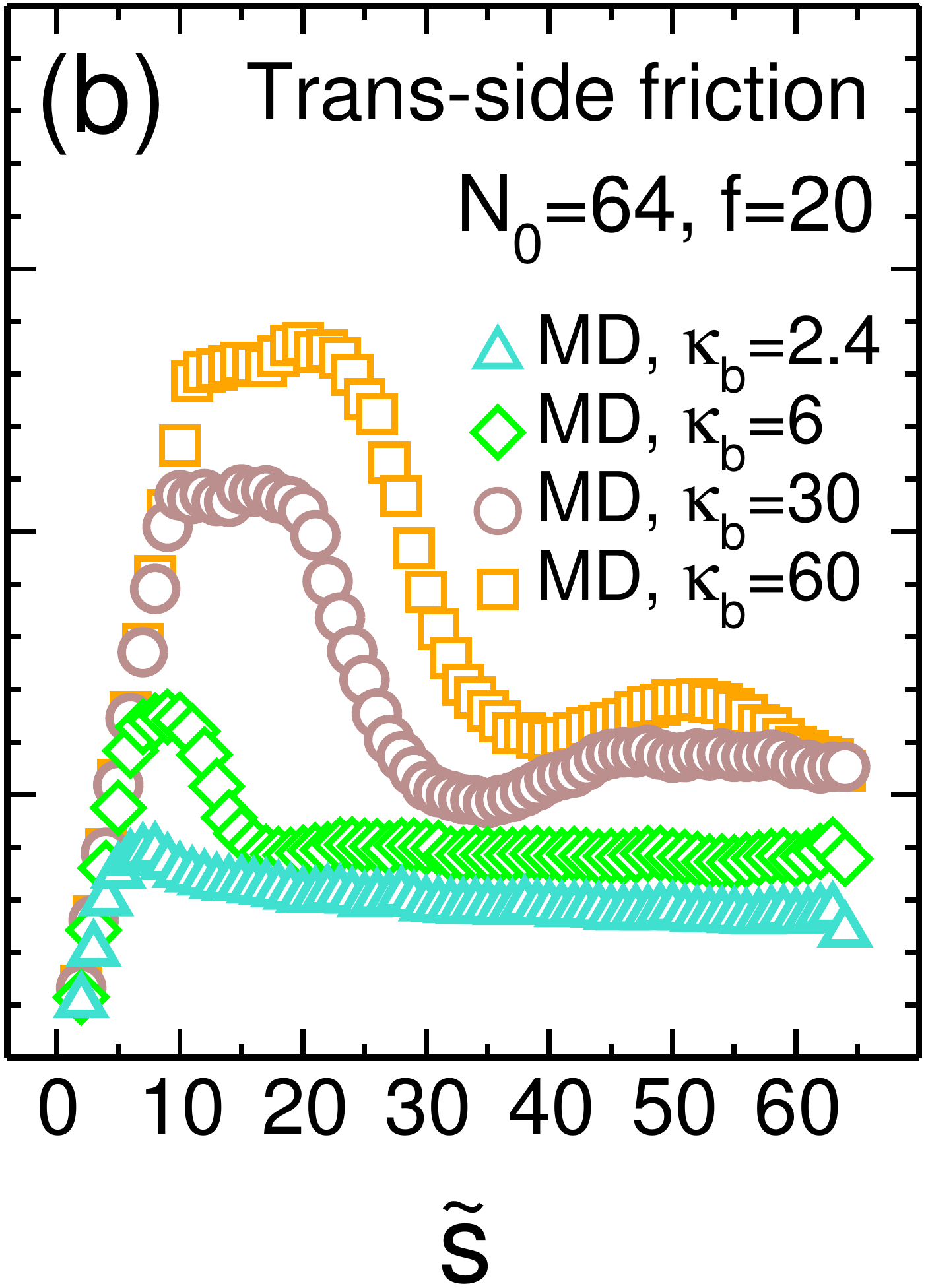}
    \end{center}\end{minipage}
\caption{(a) The {\it trans} side friction $\tilde{\eta}_{_\mathrm{TS}}$ as a function of the translocation coordinate $\tilde{s}$ for 
chain length $N_0 = 64$, bending rigidity coefficient $\kappa_{{\mathrm b}}=30$ and various values of the external 
driving force $f=5, 10$ and $20$. The orange squares ($f=5$), turquoise triangles ($f=10$) and 
brown circles ($f=20$) are MD data.
For $f=20$, the black solid line represents the {\it trans} side friction at the beginning of 
the translocation process, which is proportional to the $x$ component of the end-to-end distance. 
The horizontal black dashed line shows that the {\it trans} side friction has a constant value of 
$\approx 10.63$ during the first buckling stage. Finally, the black dashed-dotted line exhibits the {\it trans} 
side friction after the buckling has occurred, demonstrating an exponential decay
to the asymptotic value of the {\it trans} side friction, $\tilde{\eta}_{_\mathrm{TS}} (\tilde{s} \rightarrow N_0)$.
The red and blue lines represent approximate analytical fits for the {\it trans} side friction for $f=5$ and $f=10$, 
respectively. (b) $\tilde{\eta}_{_\mathrm{TS}}$ as a function of $\tilde{s}$ for 
chain length $N_0 = 64$,  external driving force $f=20$ and various values of the bending rigidity coefficient 
$\kappa_{{\mathrm b}}=2.4, 6, 30$ and 60. 
The orange squares ($\kappa_{\mathrm b}=60$), brown circles ($\kappa_{\mathrm b}=30$), green diamonds ($\kappa_{\mathrm b}=6$) and turquoise triangles 
($\kappa_{\mathrm b}=2.4$) are MD data.
} 
\label{trans-side-friction_pore_driven_flexible_static_fig}
\end{center}
\end{figure*}


\subsection{End-to-end distance of a semi-flexible chain}\label{end_to_end_distance_semi_flexible_chain}

The equation of motion for $\tilde{R} (\tilde{t})$, which is the root-mean-square of 
the end-to-end distance, i.e. $\tilde{R}_N$, can be found if an analytical form of $\tilde{R} (\tilde{t})$ for semi-flexible 
chains is known. 
To this end extensive MD simulations of bead-spring models of semi-flexible chains in 3D have been carried out \cite{jalal2017SR}. 

Figure \ref{end-to-end_pore_driven_flexible_static_fig}(a) shows the normalized end-to-end distance of a semi-flexible chain $\tilde{R}_N^2/N^{2\nu}$ as a 
function of the chain length $N$ for fixed $k_{\mathrm B} T=1.2$ and different values of the bending rigidity (in the MD simulations): 
$\kappa_{{\mathrm b}}= 6$ (red diamonds), 30 (blue circles) and 120 (green squares), which correspond to persistence lengths $\ell_{\mathrm p}= 5$ 
(red dashed-dotted line), 25 (blue solid line) and 100 (green dashed-dotted-dotted line), respectively, according to 
$\ell_{{\mathrm p}}= \kappa_{{\mathrm b}}/(k_{\mathrm B} T)$ in 3D. 
The lines come from the analytical interpolation formula
\begin{eqnarray}
\tilde{R}_N = \bigg\{ &&  + \tilde{R}^2_{\mathrm F}
- \frac{\tilde{R}^4_{\mathrm F} }{2 a_1 N^2 } \bigg[ 1- \exp \bigg( - \frac{2 a_1 N^2 }{ \tilde{R}^2_{\mathrm F} }
\bigg) \bigg] \nonumber\\
&&  + 2 \tilde{\ell}_{{\mathrm p}} N \!
- \frac{2 \tilde{\ell}_{{\mathrm p}}^{2}  }{b_1}  
\bigg[ 1 \!-\! \exp \bigg( \!\!- \frac{b_1 N }{ \tilde{\ell}_{{\mathrm p}}} \bigg) \! \bigg] \! \bigg\}^{\frac{1}{2}}.
\label{end_to_end_distance}
\end{eqnarray}
Here {$\tilde{R}_{\mathrm F}= A \tilde{\ell}_{{\mathrm p}}^{\nu_{{\mathrm p}}} N^{\nu}$,} with $\ell_{{\mathrm p}}$ as the persistence length 
and $\nu_{{\mathrm p}}= 1/(d+2)$ ($d=3$), which is the scaling form of the end-to-end distance of the chain in the limit 
$N/ \tilde \ell_{\mathrm p} \gg 1$ \cite{Nakanishi} that is correctly recovered by Eq. (\ref{end_to_end_distance}).
In the limit of $N / \tilde{\ell}_{{\mathrm p}} \ll 1$, which is the rod-like or stiff chain limit, the trivial result of $\tilde{R}_N=N$ 
is also recovered by Eq. (\ref{end_to_end_distance}). The values of the constant fitting parameters 
turn out to be $A=0.8$, $a_1=0.1$ and 
$b_1=0.9$, and $\nu=0.588$ is the 3D Flory exponent.

The main panel in Fig. \ref{end-to-end_pore_driven_flexible_static_fig}(b) is the same as in (a) but only for bending rigidity 30 
that corresponds to $\ell_{\mathrm p}= 25$. 
To present the asymptotic behavior of the end-to-end distance for this persistence length, the normalized quantity 
$\tilde{R}_N/N^{2\nu}$ has been plotted for an extended range of $N$ in the inset.
The inset presents a crossover from a rod-like chain to a Gaussian (ideal) polymer where $\nu = 1/2$. Moreover, it shows that eventually 
it correctly crosses from a Gaussian intermediate range to a self-avoiding chain at very large $N/\tilde{\ell}_{\mathrm p}$ \cite{Hsu}.


\subsection{Trans side friction}\label{friction_trans_side_semi_flexible}

For a semi-flexible chain some monomers close to the pore in the {\it trans} side contribute to the friction due to their net motion 
in the direction of the external driving force. Therefore, the {\it trans} side friction must be quantified.
As the {\it trans} side friction has complicated dependence on the physical parameters of the system we have calculated it 
numerically from the MD simulations by using the normalized cosine-correlation function \cite{jalal2017SR}.
In Fig.~\ref{trans-side-friction_pore_driven_flexible_static_fig}(a) the numerically extracted friction 
has been plotted as a function $\tilde{s}$, translocation coordinate, for fixed chain length of $N_0 = 64$, 
bending rigidity $\kappa_{\mathrm b} = 30$, which corresponds to persistence length $\ell_{\mathrm {p}} = 25$ in 3D, and for three different values of the 
external driving force $f=5, 10$ and $20$. Panel (b) is the same as (a) but for a fixed value of the driving force $f=20$ and different 
values of the bending rigidity $\kappa_{\mathrm b} = 2.4, 6, 30$ and 60.
This figure reveals three distinct regimes in $\tilde{\eta}_{\mathrm TS}(\tilde{s})$. 
In the regime of small $\tilde{s}/N_0$, the friction grows proportional to $\tilde{R}_x$ which is the amplitude of $x$ component
of the end-to-end distance. After this initial stage it saturates to almost a constant value (for example 10.63 for $f=20$), 
which indicates buckling of the {\it trans} side subchain. Then the friction approximately 
exponentially decays towards another constant value ($\approx 5.5$ for $f=20$).

It should be noted that currently we do not have any analytic formula available for $\tilde{\eta}_{\mathrm TS}$ in the strong
stretching regime considered here. For weaker driving forces, the {\it trans} side friction does not exhibit the exponentially decaying 
term as the chain has more time to relax during translocation and the polymer dynamics is slower. In this case the asymptotic value of
the {\it trans} side friction will be somewhat higher than for fast translocation \cite{Linna_2018}.


\subsection{Time evolution of the tension front}

Using the analytical form of $\tilde{R} (\tilde{t})$ in Eq. (\ref{end_to_end_distance}) together with the mass conservation 
in the TP stage $N = \tilde{l} + \tilde{s}$, where $\tilde{l} = \tilde{R}$, the tension front equation of motion
for the SS regime is derived as
\begin{equation}
\dot{\tilde{R}} (\tilde{t}) \!=\!
\frac{ \tilde{\phi} (\tilde{t}) ~ (\cal{U}+\cal{Y}) }{  2 \tilde{R} (\tilde{t}) - (\cal{U}+\cal{Y}) },
\label{evolution_of_R_propagation_SS}
\end{equation}
where 
\begin{eqnarray}
{\cal{U}} &=& +  \frac{ \tilde{R}^2_{\mathrm F} }{ N } 
\big[ 2\nu -(2-2\nu) \exp \! \big( -2 a_1 N^{2} / \tilde{R}^2_{\mathrm F} \big) \big] \nonumber\\
&&+ \frac{(4\nu-2) \tilde{R}^4_{\mathrm F} }{2 a_1 N^3 } 
\big[ -1 + \exp \! \big( -2 a_1 N^{2} / \tilde{R}^2_{\mathrm F} \big) \big], \nonumber\\
{\cal{Y}} &=& 2 \tilde{\ell}_{{\mathrm p}} \big[ 1 - \exp \! \big( - b_1N / \tilde{\ell}_{{\mathrm p}}  \big) \big].
\label{G_evolution_of_R_propagation_SS_2}
\end{eqnarray}
In the PP stage where the correct closure relation is $\tilde{l}+\tilde{s}=N_0$, 
the equation of motion for the tension front can be derived as
\begin{equation}
\dot{\tilde{R}} (\tilde{t}) = - \tilde{\phi} (\tilde{t}).
\label{evolution_of_R_post_propagation_SS}
\end{equation}
The force balance equation for the semi-flexible chain is the same as in Eq. (\ref{BD_eq_felxible-static}), 
i.e. $\tilde{\Gamma} (\tilde{t}) d \tilde{s}/{ d \tilde{t}} = \tilde{f}$, but the friction due to 
the {\it trans} side must be explicitly taken into account in the effective friction. 
The friction coefficient and the monomer flux for the semi-flexible polymer translocation are then
\begin{equation}
\tilde{\phi} (\tilde{t}) = \frac{ \tilde{f} }{ \tilde{R} (\tilde{t}) + \tilde{\eta}_{\mathrm{p}} + \tilde{\eta}_{\mathrm{TS}} } ;
\hspace{+0.2cm}
\tilde{\Gamma} (\tilde{t}) = \tilde{R} (\tilde{t}) + \tilde{\eta}_{\mathrm{p}} + \tilde{\eta}_{\mathrm{TS}}.
\label{phi_and_friction_semi-flexible}
\end{equation}
To find the solution for the model, in the TP stage Eqs. (\ref{BD_eq_felxible-static}), (\ref{evolution_of_R_propagation_SS}) 
and (\ref{phi_and_friction_semi-flexible}) must be solved self-consistently 
while in the PP stage one must solve Eqs. (\ref{BD_eq_felxible-static}), (\ref{evolution_of_R_post_propagation_SS}) 
and (\ref{phi_and_friction_semi-flexible}).

\begin{figure*}[t]\begin{center}
    \begin{minipage}[b]{0.4\textwidth}\begin{center}
        \includegraphics[width=1.0\textwidth]{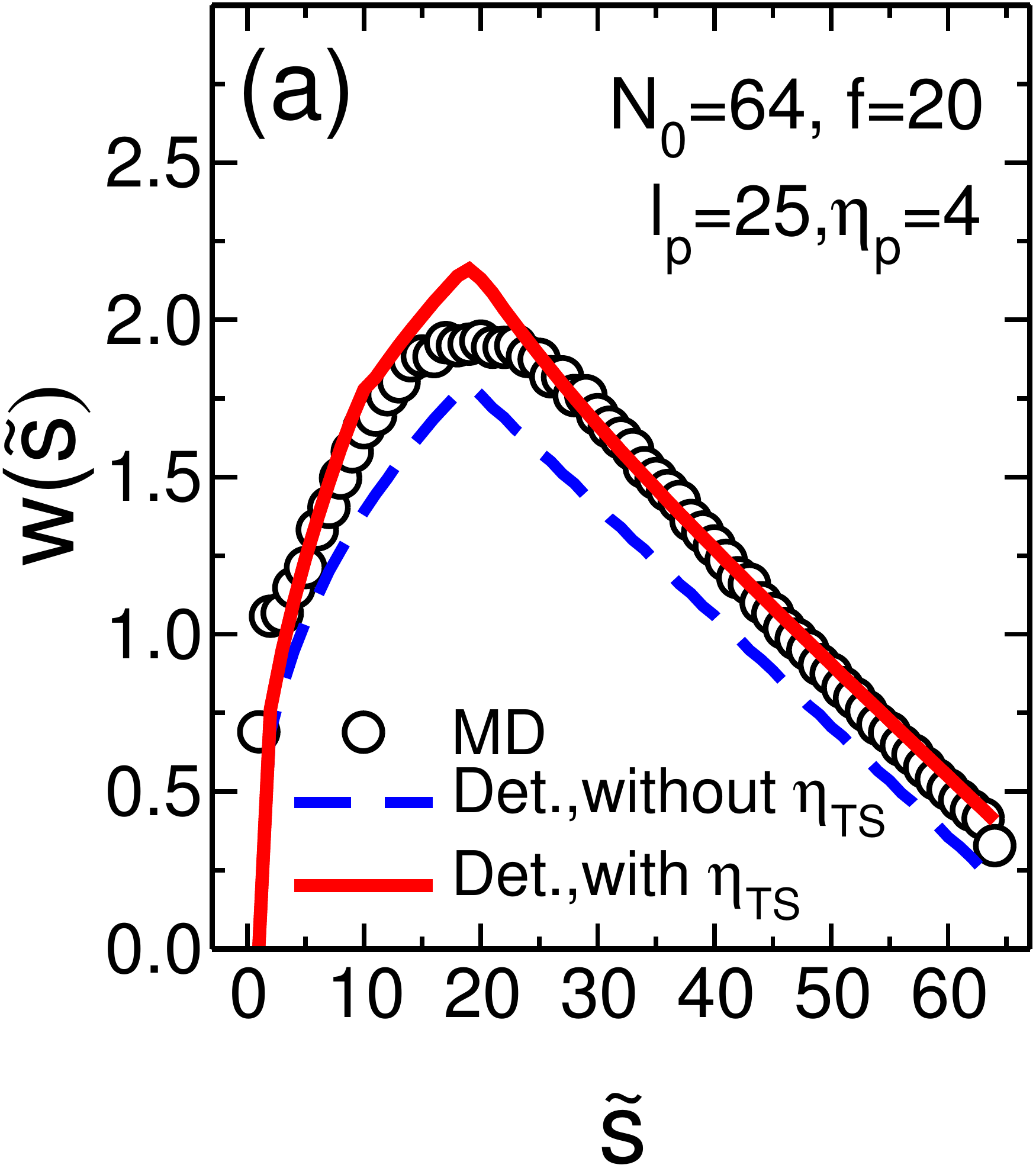}
    \end{center}\end{minipage} \hskip-0.05cm
    \begin{minipage}[b]{0.394\textwidth}\begin{center}
        \includegraphics[width=1.0\textwidth]{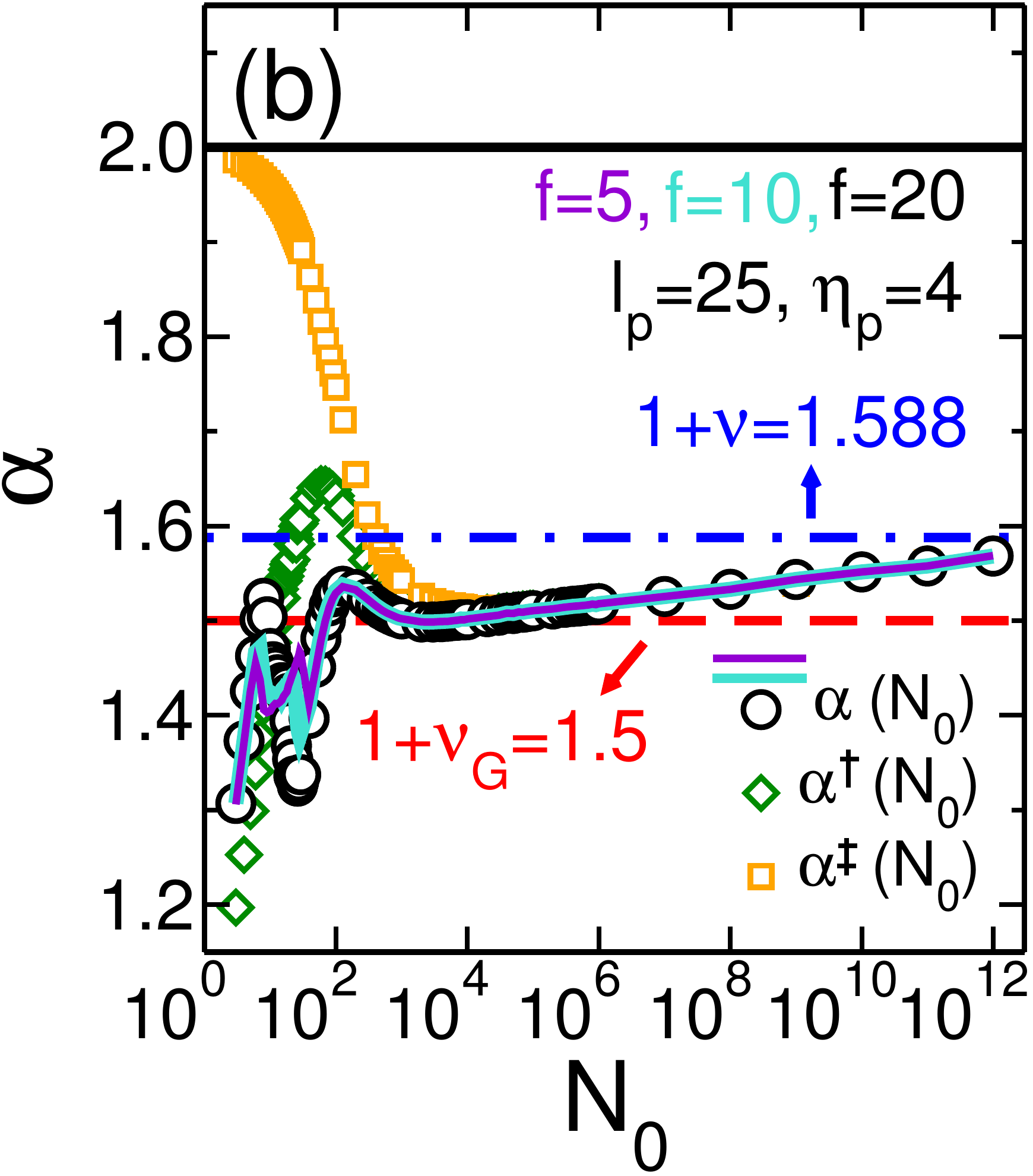}
    \end{center}\end{minipage}
\caption{(a) The WT distribution $w (\tilde{s})$ as a function of $\tilde{s}$, for a semi-flexible polymer translocation process,
with fixed values of $N_0=64$, $\ell_{{\mathrm p}}=25$, $f=20$ and $\eta_{\textrm{p}}= 4$, for the chain length, persistence length, 
the external driving force, and pore friction, respectively. No noise is added to the IFTP theory here.
The black circles are the MD simulation data. While the blue dashed line is the WT in the absence of the {\it trans} 
side friction \cite{jalal2014}, the solid red line presents the result from the IFTP theory including $\tilde{\eta}_{\mathrm TS}$.
(b) The translocation time exponent $\alpha$ as a function of $N_0$, for fixed values of
$\ell_{\mathrm{p}}=25$ and $\eta_{\textrm{p}}=4$, and for various values of the driving force $f=5$ (violet solid line), 
$10$ (turquoise solid line) and $20$ (black circles).
The rescaled translocation exponents $\alpha^{\dag}$ and $\alpha^{\ddag}$ are shown by the green diamonds and 
orange squares, respectively. The horizontal black solid, red dashed and blue dashed-dotted lines present 
the asymptotic rod-like, Gaussian and excluded volume chain limits, respectively.
} 
\label{WT_and_translocation_exponenttrans_semi_flexible_fig}
\end{center}
\end{figure*}


\subsection{Waiting time distribution} 

In Fig. \ref{WT_and_translocation_exponenttrans_semi_flexible_fig} the WT has been plotted as a function of the translocation coordinate,
for fixed values of $N_0=64$, $\ell_{{\mathrm p}}=25$, $f=20$ and $\eta_{\textrm{p}}= 4$, which are the chain length, persistence length, 
the external driving force, and pore friction, respectively. The black circles are the MD simulation data. 
The blue dashed line is the WT in the absence of the {\it trans} 
side friction \cite{jalal2014}, while the solid red line presents the result from the IFTP theory including 
$\tilde{\eta}_{\rm TS}$. As clearly shown in the figure, the {\it trans} side friction, $\tilde{\eta}_{\rm TS}(t)$, 
must be included in order to have a quantitative agreement between theory and MD data.


\subsection{Translocation time exponent}

As discussed in subsection \ref{translocation_time_felxible_static_subsection} the average translocation time 
for flexible chains under a constant driving force scales as $\tau = c_1 N_0^{\nu +1} + c_2 {\tilde{\eta}}_{\mathrm{p}} N_0 $, 
where $c_1$ and $c_2$ are constants. 
The asymptotic scaling, where $\alpha = {\nu +1}$, is caused by a significant finite-size correction due the second term,
which is the pore friction contribution \cite{ikonen2012a,ikonen2012b,ikonen2013,jalal2014,jalal2015}.
It should be mentioned that for the semi-flexible chains in the limit of large $N_0$ when $\tilde{\ell}_{\mathrm p}/N_0 \ll 1$
the asymptotics is recovered. 
Moreover, in the rod-like limit, where $\tilde{\ell}_{\mathrm p}/N_0 \gg 1$, the translocation time scales as $\tau \propto N_0^2$.
An analytical form of $\tau$ can be derived following Ref. \cite{jalal2014} as 
\footnote{Due to a technical mistake the authors of Ref. \cite{Lam_JStatPhys2015} got an 
incorrect scaling form for the translocation time.}
\begin{equation}
\tilde{\tau} = \frac{1}{\tilde{f}} \bigg[ \int_0^{N_0} \tilde{R}_N dN 
+ \tilde{\eta}_{\textrm{p}} N_0 \bigg]  + \tilde{\tau}_{\textrm{TS}},
\label{scaling_trans_time_SS_2}
\end{equation}
where $\tilde{\tau}_{\mathrm{TS}} = \big[ \int_0^{N_0}\tilde{\eta}_{_\mathrm{TS}} dN - 
\int_0^{\tilde{R}_{N_0}} ( \tilde{\eta}_{_\mathrm{TS,TP}} - \tilde{\eta}_{_\mathrm{TS,PP}} ) d\tilde{R} \big]/\tilde{f}$ 
is the contribution of the {\it trans} side to the translocation time.
The second term in $\tilde{\tau}_{\mathrm{TS}}$ is a result of the non-monotonic behavior of $\tilde{\eta}_{\textrm{TS}}$ 
in the TP ($\tilde{\eta}_{\mathrm{TS,TP}}$) as well as the PP ($\tilde{\eta}_{\mathrm{TS,PP}}$) stages \cite{jalal2017SR}.

In the rod limit a simple analytical result is obtained as
\begin{equation}
\tilde{\tau} = \frac{1}{\tilde{f}} \big[ \tilde{\eta}_{{\mathrm p}} N_0 + N_0^2 \big] ,
\label{scaling_trans_time_Rod_like_2}
\end{equation}
which reveals the well-known asymptotic result $\alpha = 2$.
Therefore, the effective exponent for semi-flexible chains will be between unity and two.

The {\it trans} side and pore friction influence the effective translocation exponent and can be
quantified by defining two rescaled translocation exponents $\alpha^{\dag}$ and $\alpha^{\ddag}$ 
as
$\tau^{\dag}= \tau - \tau_{\textrm{TS}} \sim N_0^{\alpha^{\dag}} $ and 
$\tau^{\ddag}= \tau - \tau_{\textrm{TS}} - a_2 {\tilde{\eta}}_{\textrm{p}} N_0 \sim N_0^{\alpha^{\ddag}} $, 
respectively.
As can be seen in Eq. (\ref{scaling_trans_time_SS_2}) the contributions from both the pore and the {\it trans}
side friction are important in the short and intermediate ($4 \lesssim N_0/ \tilde{\ell}_{{\rm p}} \lesssim 400$) regime.

In Fig. \ref{WT_and_translocation_exponenttrans_semi_flexible_fig}(b) the translocation time exponent 
$\alpha$ has been plotted as a function of $N_0$, for fixed values of the $\ell_{\mathrm{p}}=25$ and 
$\eta_{\textrm{p}}=4$, and for various values of the driving force $f=5$ (violet solid line), 
$10$ (turquoise solid line) and $20$ (black circles).
The rescaled translocation exponents $\alpha^{\dag}$ and $\alpha^{\ddag}$ are shown by the green diamonds and 
orange squares, respectively. The horizontal black solid, red dashed and blue dashed-dotted lines present 
the asymptotic rod-like, Gaussian and excluded volume chain limits, respectively.
As can be clearly seen, in an extended intermediate range of chain lengths
the translocation exponent is very close to $\alpha= 3/2$ which is the Gaussian value of the exponent. 
Then it slowly approaches the asymptotic value of $1+\nu=1.588$ from below. 
It should be mentioned that in order to see this crossover, a full scaling form for the end-to-end distance 
in Eq. (\ref{end_to_end_distance}) is needed.

\begin{figure*}[t]\begin{center}~\hskip-0.9cm
    \begin{minipage}[b]{0.258\textwidth}\begin{center}
        \includegraphics[width=1.12\textwidth]{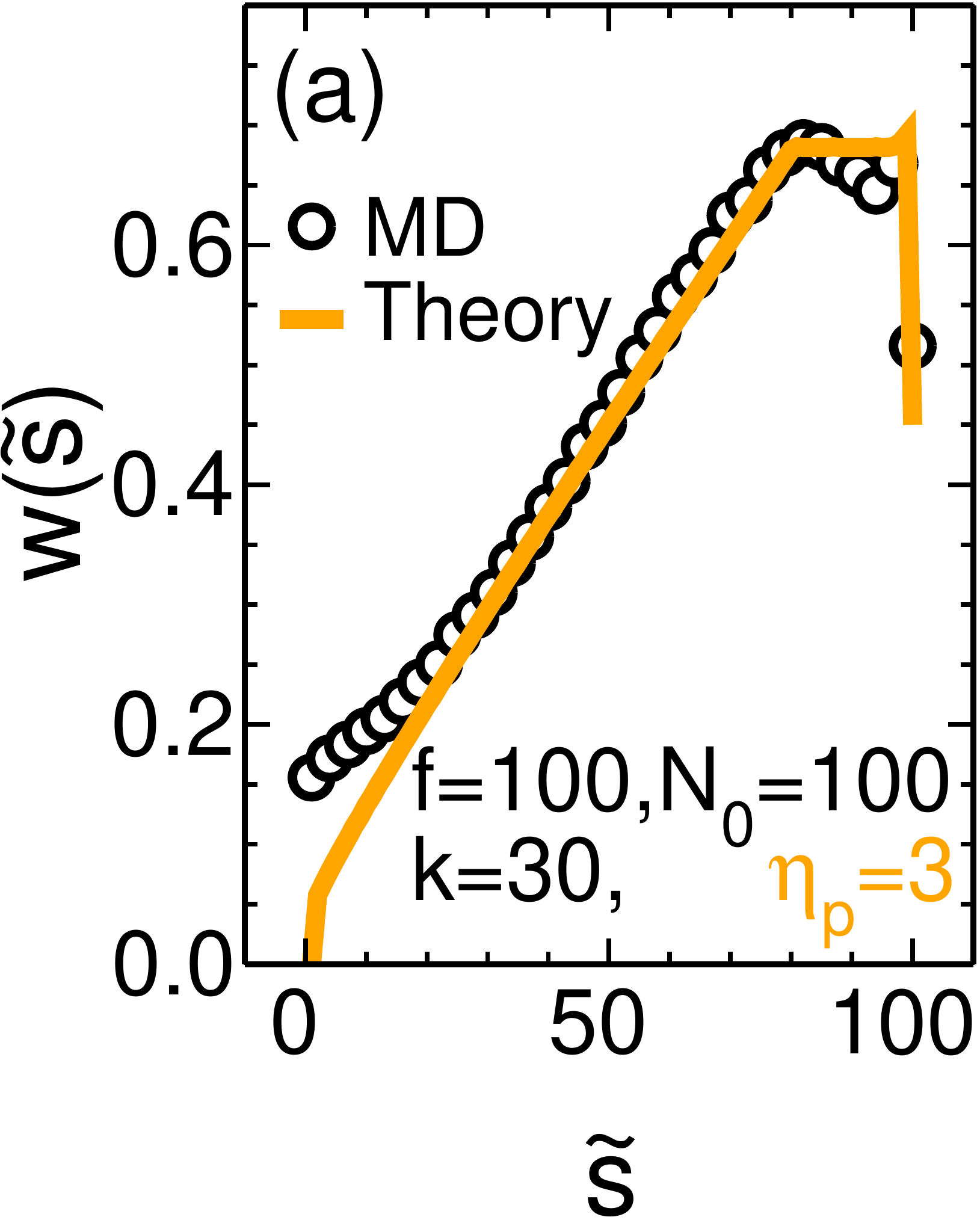}
    \end{center}\end{minipage} \hskip+0.079cm
    \begin{minipage}[b]{0.258\textwidth}\begin{center}
        \includegraphics[width=0.848\textwidth]{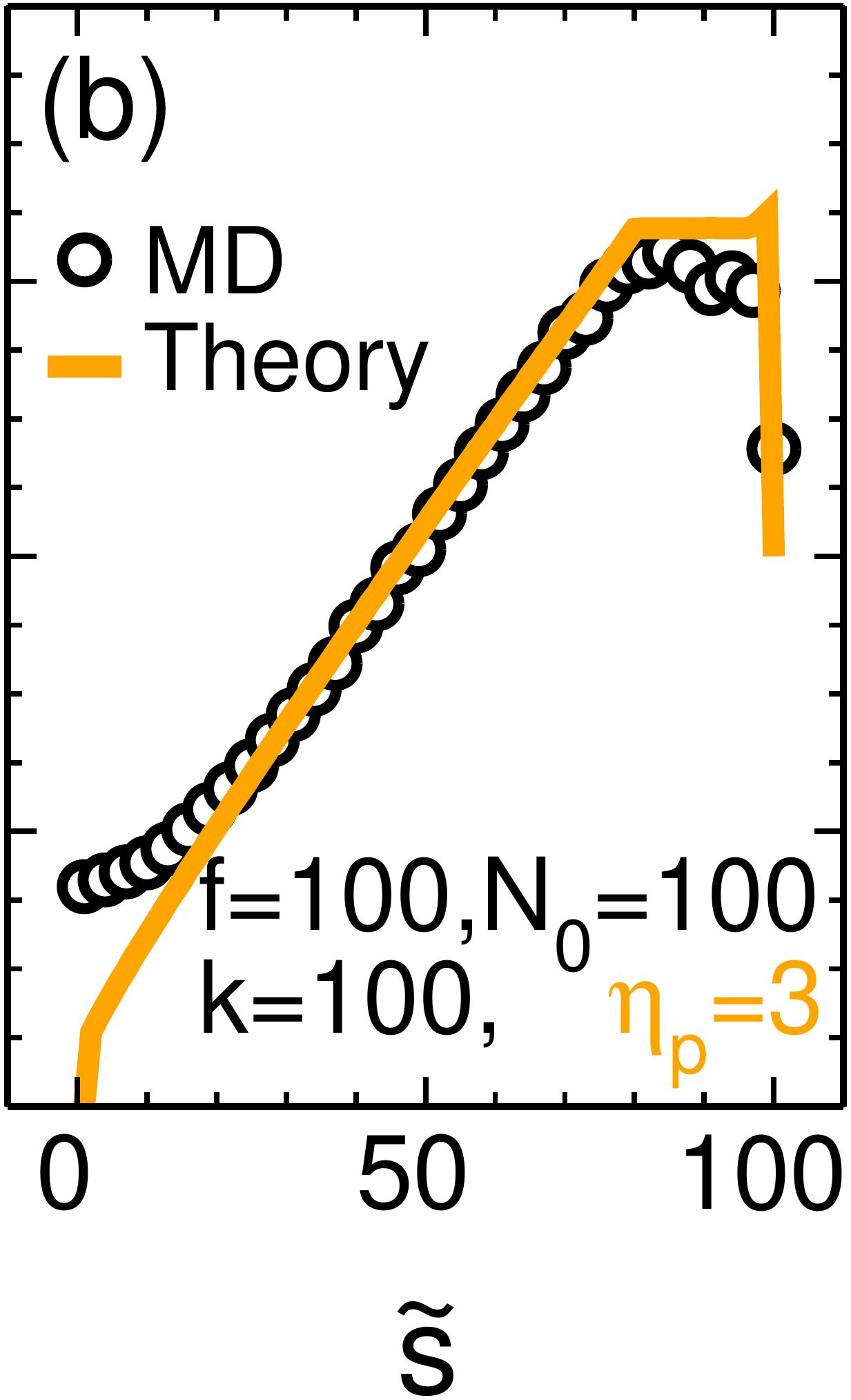}
    \end{center}\end{minipage} \hskip-0.825cm
    \begin{minipage}[b]{0.258\textwidth}\begin{center}
        \includegraphics[width=0.848\textwidth]{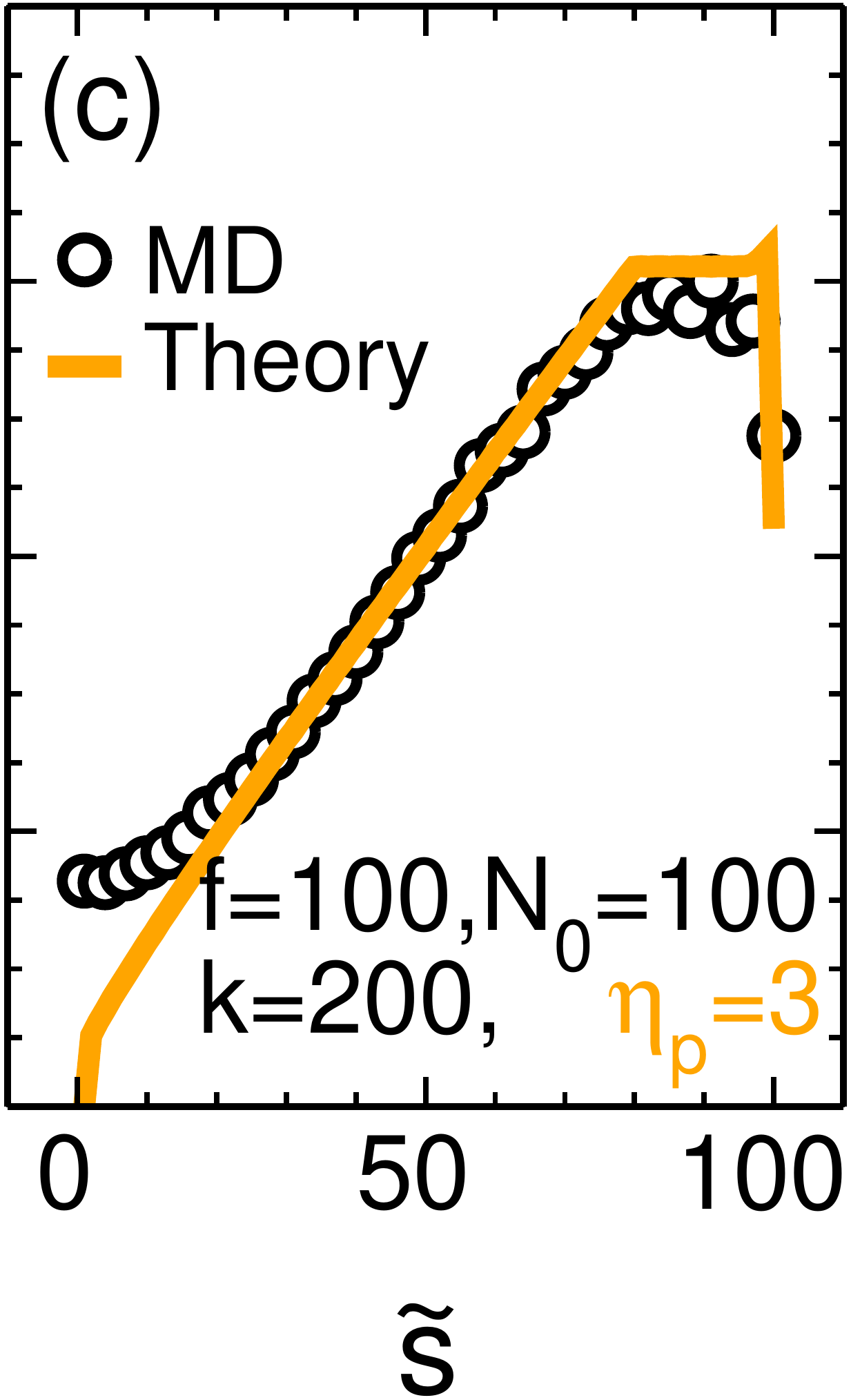}
    \end{center}\end{minipage} \hskip-0.4cm
    \begin{minipage}[b]{0.258\textwidth}\begin{center}
        \includegraphics[width=1.1\textwidth]{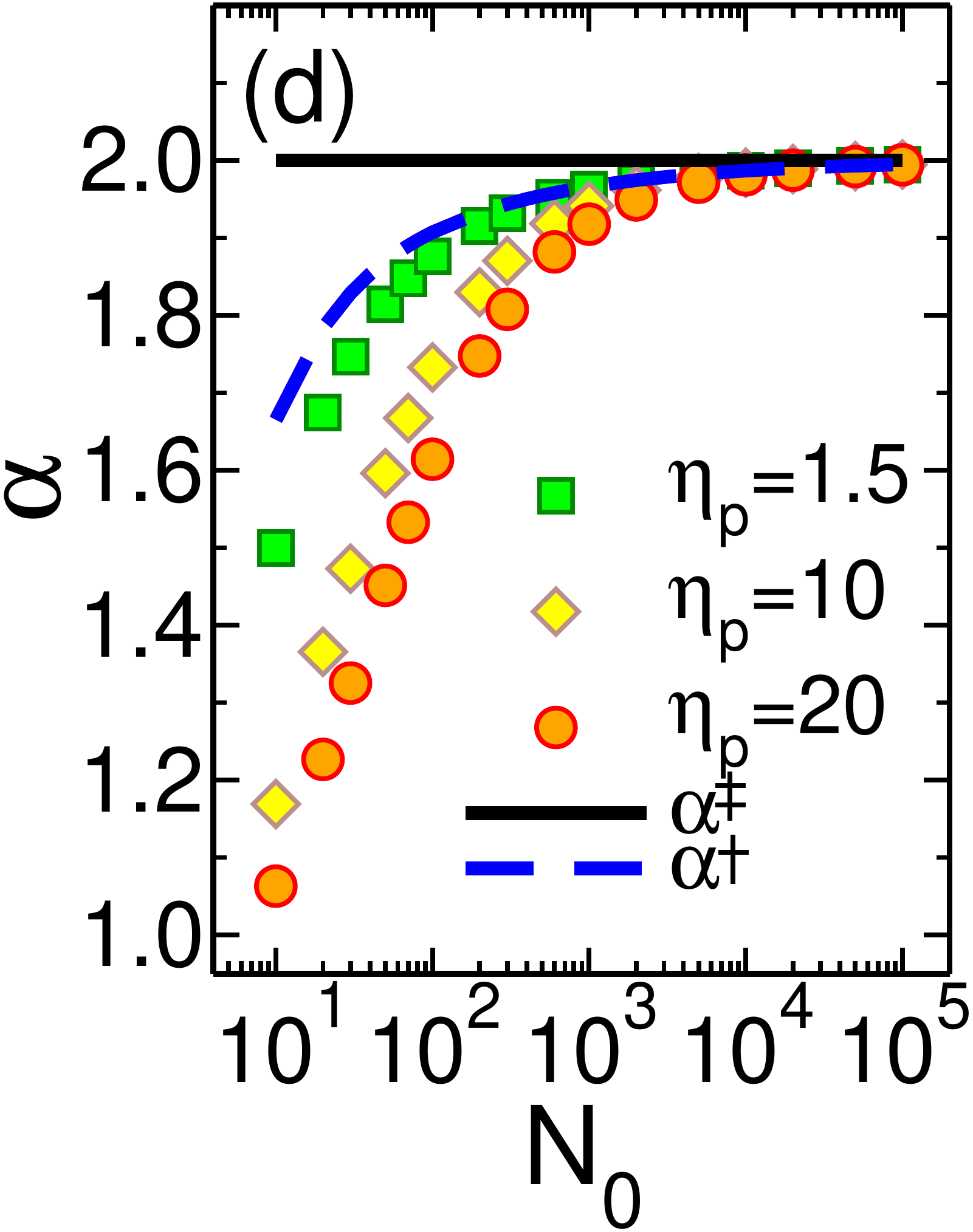}
    \end{center}\end{minipage}
\caption{(a) The WT distribution $w (\tilde{s})$ as a function of $\tilde{s}$,
for the end-pulled polymer translocation process, with constant driving force $f = 100$, chain length $N_0 = 100$, 
and pore friction $\eta_{\textrm{p}} = 3$, for the spring constant $k=30$ in the MD simulations.
The black circles present the MD simulation results while the solid orange line is the result of the IFTP theory. 
Panels (b) and (c) are the same as (a) but for different values of the spring constant $k = 100$ and 200, respectively.
(d) The effective translocation time exponent $\alpha$ as a function of $N_0$, for different values of 
the pore friction $\eta_{\mathrm{p}}=1.5$ (green squares), 10 (yellow diamonds) and 20 (orange circles). 
The rescaled translocation exponents $\alpha^{\dag}$ and $\alpha^{\ddag}$ are shown by the blue dashed 
and the horizontal black solid lines, respectively. See text for details.
} 
\label{WT_trans_time_exponent_end_pulled_flexible_fig}
\end{center}
\end{figure*}
%


\section{End-pulled flexible chain} \label{end-pulled_flexible_static-pore_constant-force}

This section is devoted to the dynamics of end-pulled polymer translocation through a nanopore,
where the external driving force only acts on the head monomer of the chain on the {\it trans} side 
(see Figs. \ref{fig:schimatic}(c) and (d)). 
To this end we generalize the IFTP theory to include the {\it trans} side subchain friction. 
Depending on the configurations of the subchain in the {\it cis} and the {\it trans} side
a complicated scenario of multiple scaling regimes is revealed.
In the high driving force limit, where the {\it trans} side subchain is strongly stretched (the SS regime), 
the theory is in excellent agreement with MD simulations. In the SS regime an exact analytical form for the 
translocation time can be derived as a function of the chain length and the external force. 
Moreover, the scaling exponents for ${\tau} \sim N_0^{\alpha} f^{\beta}$ in the asymptotics
are $\alpha=2$, and $\beta =-1$. 
The correction-to-scaling terms arising due to the {\it cis} side and pore friction are revealed by the IFTP theory.
These terms lead to a very slow approach to the asymptotic exponent $\alpha =2$ from below as a function
of increasing chain length $N_0$.

In the SS regime when the {\it cis} side subchain is also in the strong stretching regime (SSC), 
the time evolution of the tension front in the TP and PP stages are given by Eqs. (\ref{evolution_of_R_TP_SS_static}) 
and (\ref{evolution_of_R_PP_SS_static}), respectively \cite{jalal2017EPL}. The evolution of the monomer flux, 
and the total effective friction, are expressed by 
$\tilde{\phi} (\tilde{t}) = \tilde{f} / [ \tilde{R} (\tilde{t}) + \tilde{\eta}_{\mathrm{p}} + \tilde{\eta}_{\mathrm{TS}} ] $, 
and 
$\tilde{\Gamma} (\tilde{t}) = \tilde{R} (\tilde{t}) + \tilde{\eta}_{\mathrm{p}} + \tilde{\eta}_{\mathrm{TS}}$, respectively, 
which are similar to those in Eq. (\ref{phi_and_friction_semi-flexible}). However, there's an important difference as compared to
the semi-flexible pore-driven case. The {\it trans} side friction for 
the end-pulled case in the SS regime, where the {\it trans} subchain is fully straightened,
is given analytically by $\tilde{\eta}_{\textrm{TS}} = \tilde{s}$ \cite{jalal2017EPL}. It should be noted that in the SS regime, 
the {\it cis} side subchain configuration may be in either the trumpet (TRC), stem-flower (SFC) or SSC regimes.
Here we only consider the SSC regime. The form of the time evolution of the tension front in 
the SFC and TRC regimes have been explained in detail in Ref. \cite{jalal2017EPL}. The equations for the time evolution 
of the monomer flux and the total effective friction in the SFC and TRC regimes are the same as of the SSC regime.


\subsection{Waiting time distribution} 

To test the validity of the IFTP theory for the end-pulled case in Fig. \ref{WT_trans_time_exponent_end_pulled_flexible_fig}(a)
we present the monomer WT distribution as a function of $\tilde{s}$, 
with constant driving force $f = 100$, chain length $N_0 = 100$, and $\eta_{\textrm{p}} = 3$ (pore friction in the theory), 
for the spring constant $k=30$ in the MD simulations. 
The black circles present the MD simulation results while the solid orange line 
is the result of the IFTP theory. Panels (b) and (c) are the same as (a) but for different values of the spring constant $k = 100$ and 200, 
respectively. 
The orange solid lines come from the IFTP theory when we solve the equations of motion with a 
combination of all the three SSC, SFC and TRC regimes (a full description for the SFC and TRC regimes can be found in Ref. \cite{jalal2017EPL}). 
The equations are solved in the SSC, or SFC or TRC regime if $\tilde{f}_0 \gtrsim N_0 -\tilde{s}$, or $ 1 \lesssim \tilde{f}_0 < N_0 - \tilde{s}$ or
$\tilde{f}_0 \lesssim 1$, respectively. $\tilde{f}_0$ has been introduced in Eq. (\ref{f0}), which for the end-pulled case under a constant driving force
is $\tilde{f}_0 \equiv \tilde{f} - \tilde{\eta}_{\mathrm{p}} \tilde{\phi} (\tilde{t}) - \tilde{\eta}_{\mathrm{TS}} \tilde{\phi} (\tilde{t}) $.
As can be seen in Figs. \ref{WT_trans_time_exponent_end_pulled_flexible_fig}(a)-(c) the IFTP theory result underestimates 
the WT for small values of $\tilde{s}$. This happens due to the stretching of the bonds and also because of the beginning of the mobile part 
reorientation on the {\it cis} side. This discrepancy occurs only for small values of $\tilde{s}$, and as the larger values of $\tilde{s}$ 
(over the whole PP stage and almost at the end of the TP stage) mainly contribute to the WT and consequently to the translocation time,
the overall behavior of the translocation process is faithfully predicted by the IFTP theory.


\subsection{Scaling exponents for translocation} 

Similar to the previous sections and following Refs. \cite{jalal2014} and \cite{jalal2017SR} an exact analytic expression for $\tilde{\tau}$
can be derived as
\begin{equation}
\tilde{\tau} = \frac{1}{\tilde{f}} \big[ \int_0^{N_0} \tilde{R}_N dN + \tilde{\eta}_{\mathrm{p}} N_0 \big] + \tilde{\tau}_{\mathrm{TS}},
\label{scaling_trans_time_1}
\end{equation}
where 
\begin{equation}
\tilde{\tau}_{\mathrm{TS}}
= \frac{N_0^2}{2\tilde{f}},
\label{scaling_trans_time_2}
\end{equation}
is the {\it trans} side friction contribution to the translocation time, due to the fully straightened {\it trans} side mobile subchain 
(see Figs. \ref{fig:schimatic}(c) and (d)). Therefore,
\begin{equation}
\tilde{\tau} = \frac{1}{\tilde{f}} \bigg[ \frac{A_{\nu}}{\nu + 1} N_0^{\nu +1} + \tilde{\eta}_{\mathrm{p}} N_0 + \frac{1}{2} N_0^2 \bigg],
\label{scaling_trans_time}
\end{equation}
where the force exponent is thus $\beta = -1$.
It should be mentioned that in the right hand side of Eq. (\ref{scaling_trans_time}), the first two terms are identical to 
the pore-driven flexible chain case \cite{jalal2014}, and the new third term, which is proportional to $N_0^2$, is due to 
the explicit form of the {\it trans} side friction, $\tilde{\eta}_{\mathrm{TS}} = \tilde{s}$ for the end-pulled case when 
the {\it trans} side is fully straightened.
As is clear in Eq. (\ref{scaling_trans_time}) the asymptotic translocation exponent is $\alpha=2$, and there are two 
correction-to-asymptotic-scaling terms which lead to pronounced crossover behavior due to the contributions of the {\it cis} 
side and pore friction to the total effective friction. 
In Fig. \ref{WT_trans_time_exponent_end_pulled_flexible_fig}(d), the translocation exponents $\alpha$, $\alpha^{\dag}$, and $\alpha^{\ddag}$
are plotted as a function of the chain length $N_0$ for different values of the pore friction, $\eta_{\mathrm{p}}= 1.5, 10$ and 20,
where the last two rescaled exponents are defined as 
\begin{eqnarray}
\tilde{\tau}^{\dag}  &=& \tilde{\tau} - \tilde{\eta}_{\mathrm{p}} N_0 / \tilde{f} \sim N_0^{\alpha^\dag}, \nonumber\\
\tilde{\tau}^{\ddag} &=& \tilde{\tau}   - \frac{1}{\tilde{f}} \big[ \int_{0}^{N_0} \tilde{R}_N dN + \tilde{\eta}_{\mathrm{p}} N_0 \big]   \sim N_0^{\alpha^{\ddag}}.
\label{tau_rescaled}
\end{eqnarray}
As Fig. \ref{WT_trans_time_exponent_end_pulled_flexible_fig}(d) clearly shows for typical parameters used here and
in most computer simulations, the asymptotic scaling is recovered
for very long chains only.


\section{Pore-driven flexible chain with a flickering pore and an oscillating force} \label{pore-driven_flexible_flickering-pore_alternating-force}

\begin{figure*}[t]\begin{center}
    \begin{minipage}[t]{0.5075\textwidth}\begin{center}
        \includegraphics[width=1.0\textwidth]{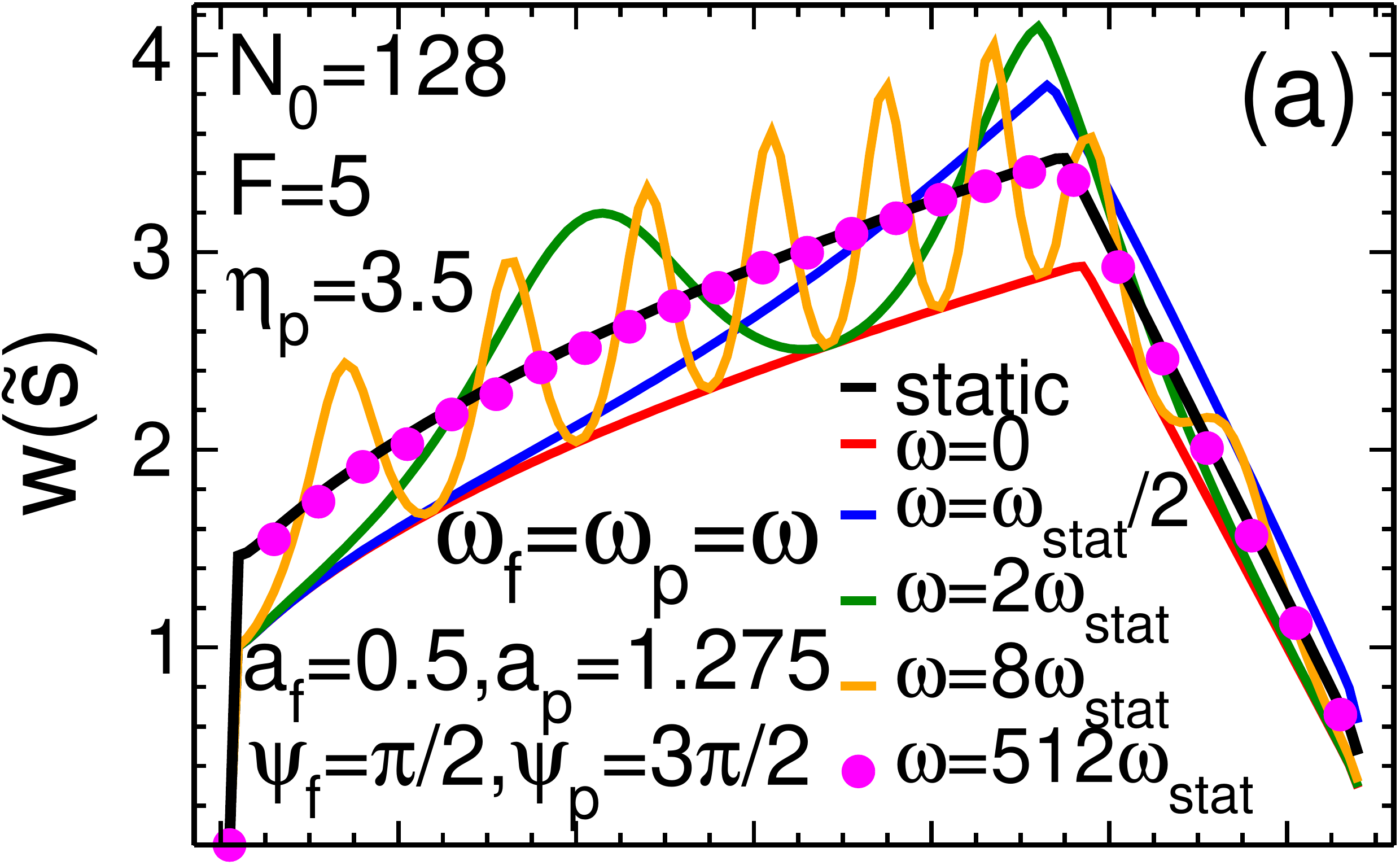}
    \end{center}\end{minipage} \hskip-0.129cm
    \begin{minipage}[t]{0.44\textwidth}\begin{center}
        \includegraphics[width=1.0\textwidth]{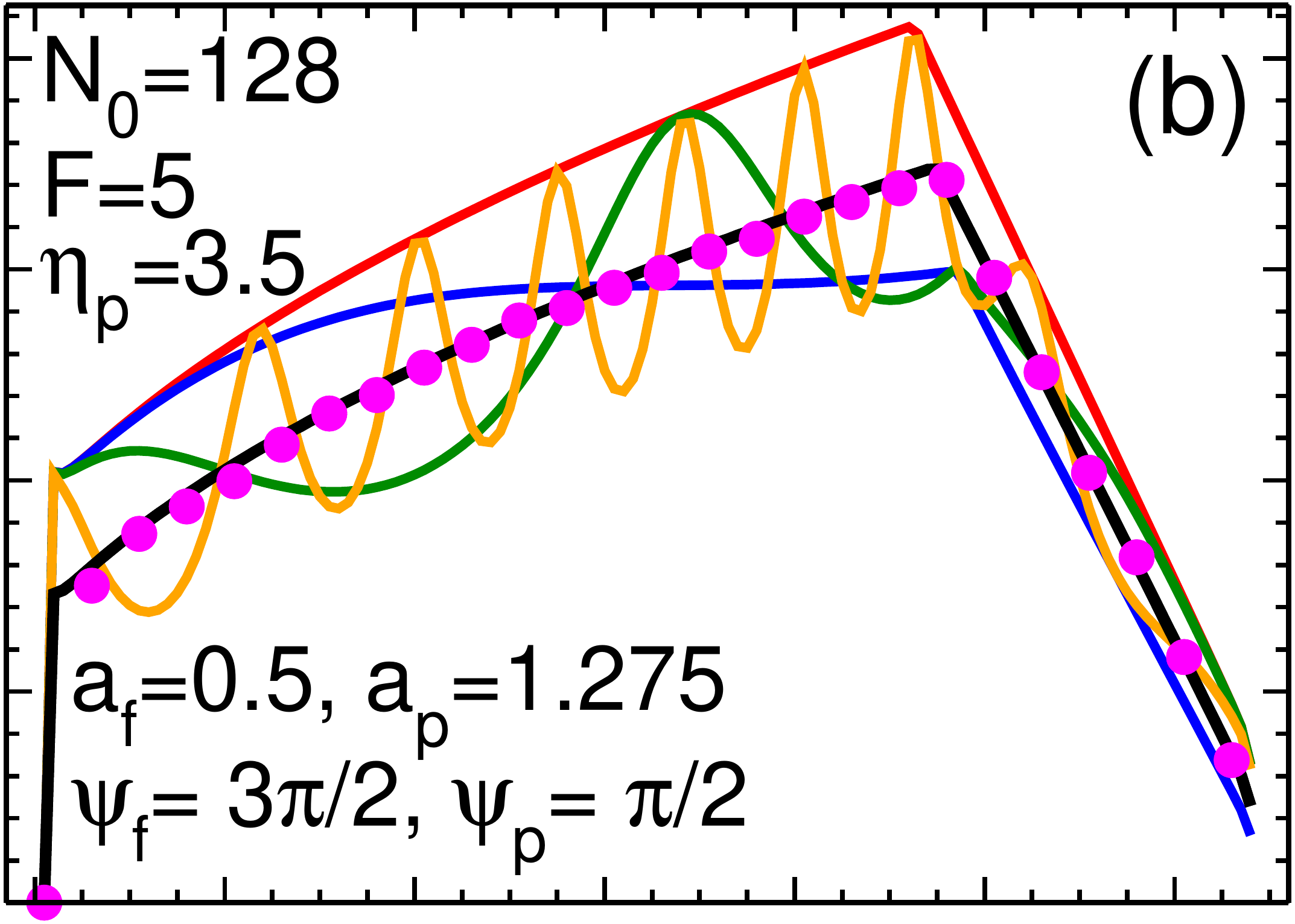}
    \end{center}\end{minipage} \vskip-0.14cm ~\hskip-0.1cm 
    \begin{minipage}[b]{0.44\textwidth}\begin{center}
        \includegraphics[width=1.153\textwidth]{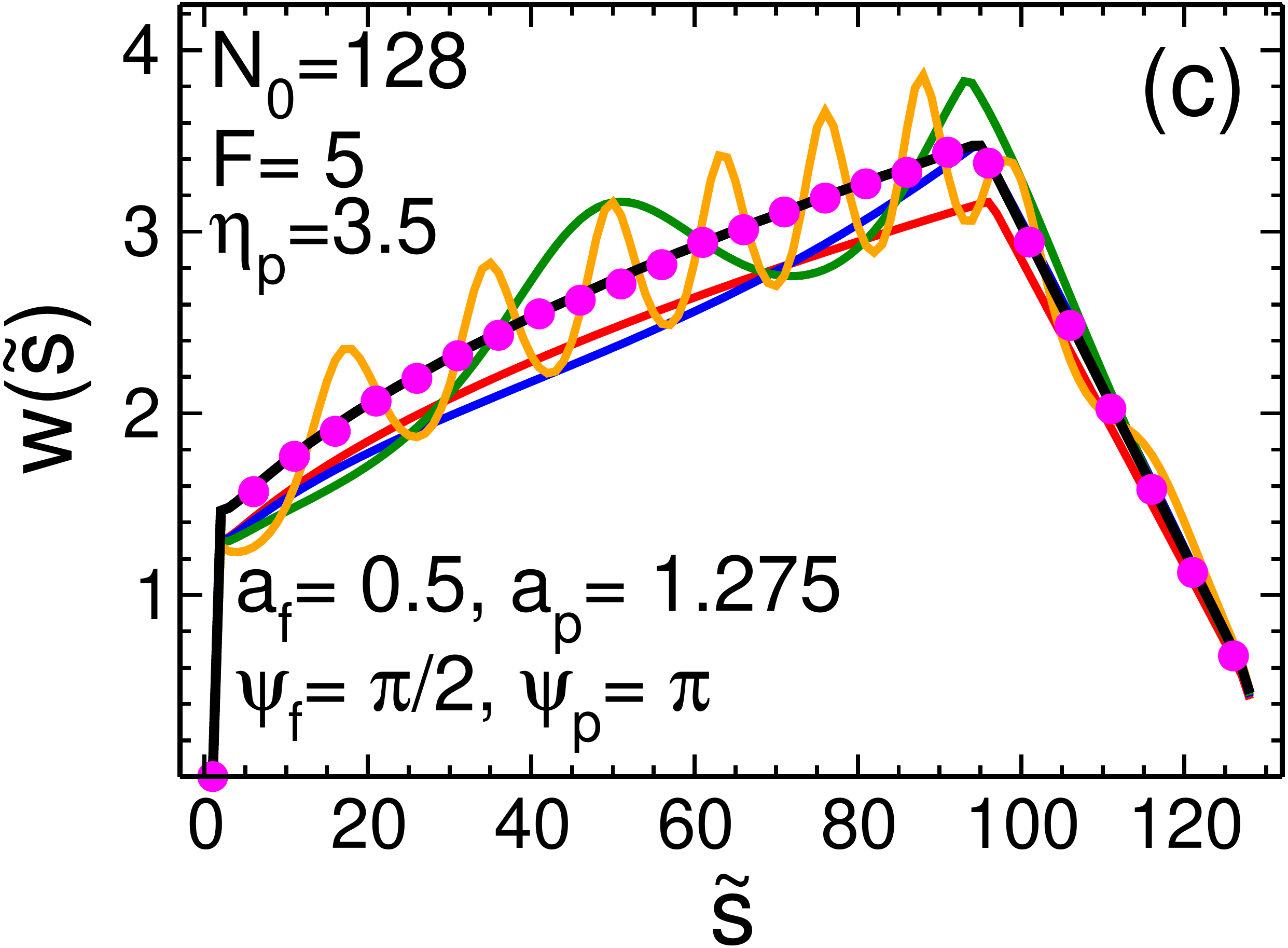}
    \end{center}\end{minipage} \hskip+1.08cm
    \begin{minipage}[b]{0.44\textwidth}\begin{center}
        \includegraphics[width=1.0\textwidth]{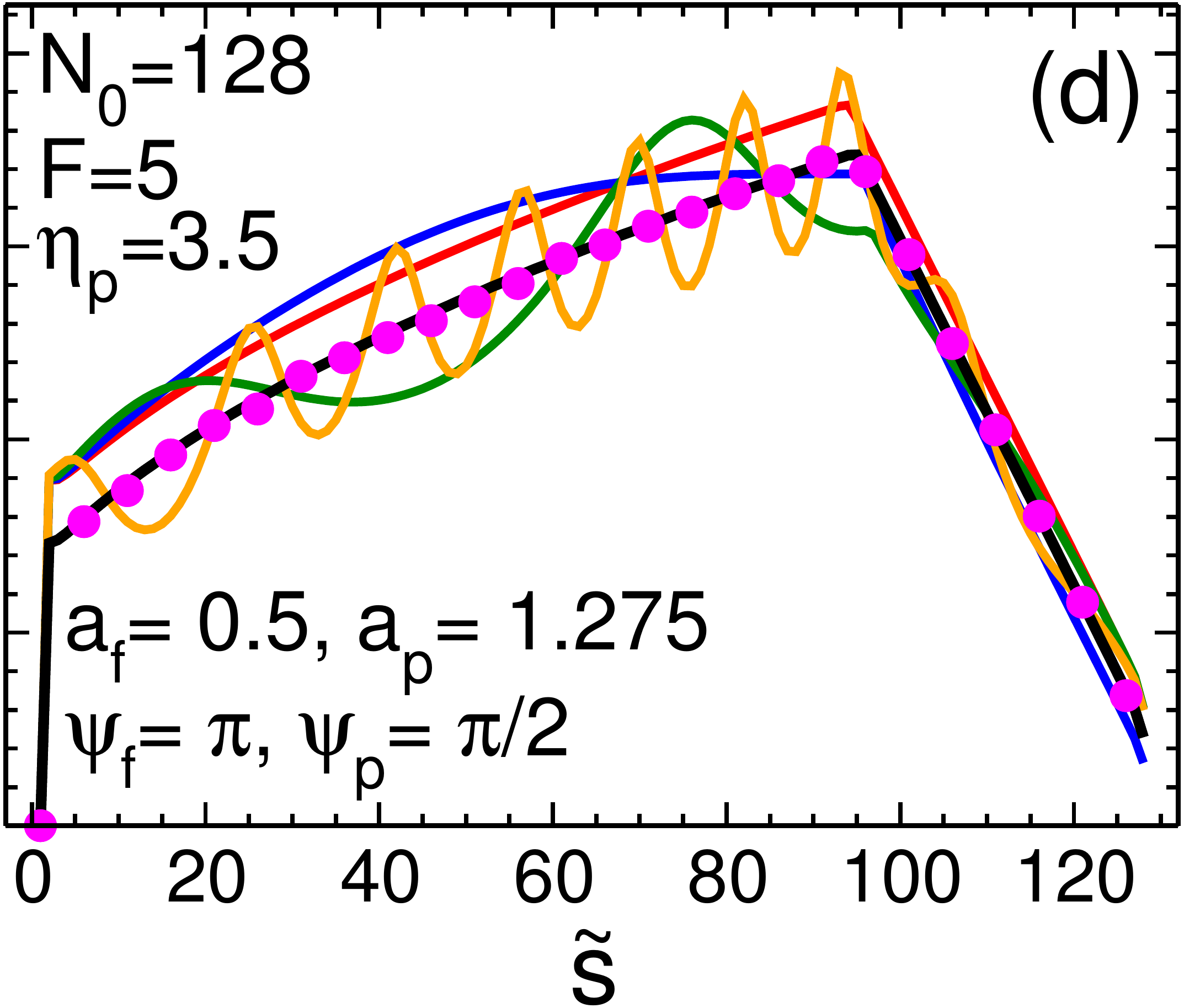}
    \end{center}\end{minipage}
\caption{(a) The waiting time $w(\tilde{s})$ as a function of the translocation coordinate $\tilde{s}$ 
for pore driven translocation of a flexible chain through a flickering pore and under a periodically oscillating driving force, 
with fixed values of the initial phases of the force $\psi_f = \pi/2$ and of the pore $3\pi/2$,
for the static pore and also for various values of the force and pore frequencies $\omega_{f} = \omega_{\mathrm{p}}= 0, \omega_{\mathrm{stat}}/2, 
2\omega_{\mathrm{stat}}, 8 \omega_{\mathrm{stat}}$ and $512\omega_{\mathrm{stat}}$, where we have defined
$\omega_{\mathrm{stat}}= 2\pi /\tau_{\mathrm{stat}}$ 
through the average translocation time $\tau_{\mathrm stat}$ for a static pore under a constant driving force. 
Here, the flickering pore friction is given by 
$\tilde{\eta}_{\mathrm{p}} (\tilde{t}) = \tilde{\eta}_{\mathrm{p}} + \tilde{a}_{\mathrm{p}} \sin (\tilde{\omega}_{\mathrm{p}} \tilde{t} +\psi_{\mathrm{p}} )$,
where $\eta_{\mathrm{p}}=3.5$ and $a_{\mathrm{p}} = 1.275$,
the oscillating external driving force is 
$\tilde{F} (\tilde{t}) = \tilde{f} + \tilde{a}_{f} \sin (\tilde{\omega}_f \tilde{t} +\psi_f )$, where $f = 5$ and $a_f = 0.5$.
Panels (b), (c) and (d) are the same as panel (a) but for different values of the initial force and pore phases
$\psi_f= 3\pi/2$ and $\psi_{\mathrm{p}}=\pi/2$, $\psi_f= \pi/2$ and $\psi_{\mathrm{p}}=\pi$,
and $\psi_f= \pi$ and $\psi_{\mathrm{p}}=\pi/2$, respectively.
} 
\label{WT_trans_time_pore_driven_flexible_flickering_fig}
\end{center}
\end{figure*}

In this section we consider pore-driven polymer translocation under an alternating driving force through a flickering pore. 
Here the alternating driving force is assumed to be periodic, and it can be directly incorporated into the IFTP theory. 
Therefore, the form of the force balance equation 
Eq.~(\ref{BD_equation}) remains the same, and $\tilde{\Gamma} (\tilde{t}) (d \tilde{s} / d \tilde{t}) = \tilde{F} (\tilde{t})$.
The flickering pore is modeled by assuming that the pore friction has a time dependent form $\tilde{\eta}_{\textrm{p}} (\tilde{t})$
and again we assume for simplicity that flickering is periodic.
Following Sec. \ref{Sec_pore-driven_flexible_static-pore_constant-force} for a fully flexible self-avoiding polymer 
the dynamical contribution of the {\it trans} side friction is insignificant~\cite{ikonen2012a,ikonen2012b,ikonen2013,dubbeldam2014}
and can be absorbed into the pore friction \cite{jalal2014,jalal2015}. Therefore, the total effective friction $\tilde{\Gamma} (\tilde{t})$ 
has both the {\it cis} side contribution $\tilde{\eta}_{\rm cis} (\tilde{t})= \tilde{R} (\tilde{t}) $ and the time dependent pore friction, 
and can be written as $\tilde{\Gamma} (\tilde{t}) = \tilde{R} (\tilde{t}) + \tilde{\eta}_{\textrm{p}} (\tilde{t})$.
Following derivations in the previous sections the flux of monomers $\tilde{\phi} (\tilde{t})$ and the effective friction $\tilde{\Gamma} (\tilde{t})$ 
are then obtained as
\begin{equation}
\tilde{\phi} (\tilde{t}) = \frac{ \tilde{f} (\tilde{t}) } { \tilde{R} (\tilde{t}) + \tilde{\eta}_{\textrm{p}} (\tilde{t}) }; 
\hspace{+1.0cm} 
\tilde{\Gamma} (\tilde{t}) = \tilde{R}(\tilde{t}) + \tilde{\eta}_{\mathrm{p}} (\tilde{t}) . 
\label{Phi_Gamma_flexible_flickering}
\end{equation}
For the SS regime
one can show again that the time evolution of the tension front in the TP and PP stages remains the same as in Eqs. (\ref{evolution_of_R_TP_SS_static}) and 
(\ref{evolution_of_R_PP_SS_static}), respectively. 
Therefore, the self-consistent solution for the IFTP theory in the TP stage is obtained from Eqs. (\ref{BD_equation}),
(\ref{evolution_of_R_TP_SS_static}) and (\ref{Phi_Gamma_flexible_flickering}), and in the PP stage the set 
of Eqs. (\ref{BD_equation}), (\ref{evolution_of_R_PP_SS_static}) and (\ref{Phi_Gamma_flexible_flickering}) must be solved.


\subsection{Waiting time} \label{waiting_time_felxible_flickering}

To proceed further we choose the driving force as a combination of 
a constant force component $\tilde{f}$ and an oscillatory term $\tilde{A}_{f} (\tilde{t})$ as
\begin{equation}
\tilde{F} (\tilde{t}) = \tilde{f} + \tilde{\mathcal{A}}_f (\tilde{t}),
\label{oscillatory_force}
\end{equation}
where $\tilde{\mathcal{A}}_f (\tilde{t}) = \tilde{a}_{f} \sin (\tilde{\omega}_{f} \tilde{t} +\psi_{f} )$,
and $\tilde{a}_{f}$, $\psi_{f}$ and $\tilde{\omega}_{f}$ are the amplitude, initial phase and frequency of the force, respectively.
Similarly, we use a simple periodic function for the pore friction coefficient to model the flickering pore as
\begin{equation}
\tilde{\eta}_{\textrm{p}} (\tilde{t}) = \tilde{\eta}_{\textrm{p}} + \tilde{\mathcal{A}}_{\textrm{p}} (\tilde{t}),
\label{flickering_pore}
\end{equation}
where $\tilde{\mathcal{A}}_{\textrm{p}} (\tilde{t}) = \tilde{a}_{\textrm{p}} \sin (\tilde{\omega}_{\textrm{p}} \tilde{t} +\psi_{\textrm{p}} )$
and $\tilde{a}_{\textrm{p}}$, $\psi_{\textrm{p}}$ and $\tilde{\omega}_{\textrm{p}}$  are the amplitude, initial phase and frequency of the 
pore friction, respectively. 

As the WT accurately reveals the dynamics of the translocation process, in Fig. \ref{WT_trans_time_pore_driven_flexible_flickering_fig}(a)
we plot the WT as a function of the translocation coordinate $\tilde{s}$ 
with fixed values of the initial phases of the force $\psi_f = \pi/2$, and of the pore $\psi_{\mathrm{p}} = 3\pi/2$. The data are
for the static pore as well as for various values of the force and pore frequencies $\omega_f = \omega_{\mathrm{p}}= 0, \omega_{\mathrm{stat}}/2, 
2\omega_{\mathrm{stat}}, 8 \omega_{\mathrm{stat}}$ and $512\omega_{\mathrm{stat}}$, where $\omega_{\mathrm{stat}}= 2\pi /\tau_{\mathrm{stat}}$ 
and the subscript stat stands for static pore under a constant driving force which was explained in detail in 
Section \ref{Sec_pore-driven_flexible_static-pore_constant-force}. Here, the flickering pore friction is given by 
$\tilde{\eta}_{\mathrm{p}} (\tilde{t}) = \tilde{\eta}_{\mathrm{p}} + \tilde{a}_{\mathrm{p}} \sin (\tilde{\omega}_{\mathrm{p}} \tilde{t} +\psi_{\mathrm{p}} )$,
where $\eta_{\mathrm{p}}=3.5$ and $a_{\mathrm{p}} = 1.275$,
while the alternating external driving force is 
$\tilde{F} (\tilde{t}) = \tilde{f} + \tilde{a}_{f} \sin (\tilde{\omega}_f \tilde{t} +\psi_f )$, where $f = 5$ and $a_f = 0.5$.
Panels (b), (c) and (d) are the same as panel (a) but for different values of the initial force and pore phases
$\psi_f= 3\pi/2$ and $\psi_{\mathrm{p}}=\pi/2$, $\psi_f= \pi/2$ and $\psi_{\mathrm{p}}=\pi$,
and $\psi_f= \pi$ and $\psi_{\mathrm{p}}=\pi/2$, respectively.
The number of oscillations in the WT curves is almost given by $\omega_f / \omega_{\mathrm{stat}}$ or by $ \omega_{\mathrm{p}} / \omega_{\mathrm{stat}} $.
It is clear that as the force and pore frequencies $\omega_f = \omega_{\mathrm{p}}= 512\omega_{\mathrm{stat}}$ are in the high frequency limit, 
the WT curves for the static case (black solid line) and for the high  frequencies (pink circles) collapse.
This happens because the dynamics of the driving force and also the flickering pore is so fast that when the monomers of the polymer 
pass the pore, they only experience the average value of the fluctuating driving force as well as the average value of the pore friction.

\begin{figure*}[t]\begin{center}
    \begin{minipage}[b]{0.43\textwidth}\begin{center}
        \includegraphics[width=1.045\textwidth]{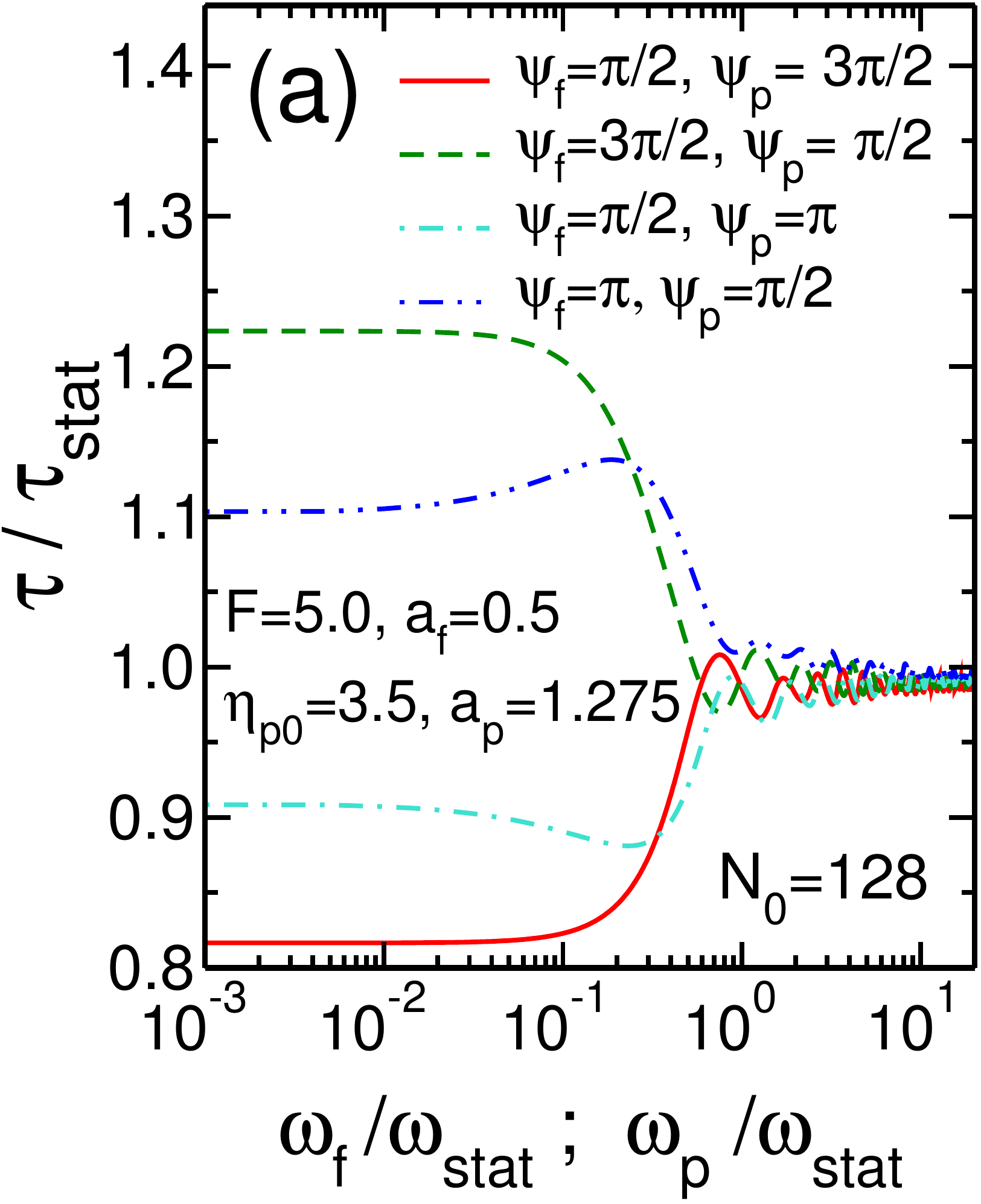}
    \end{center}\end{minipage} \hskip-0.42cm
    \begin{minipage}[b]{0.43\textwidth}\begin{center}
        \includegraphics[width=0.873\textwidth]{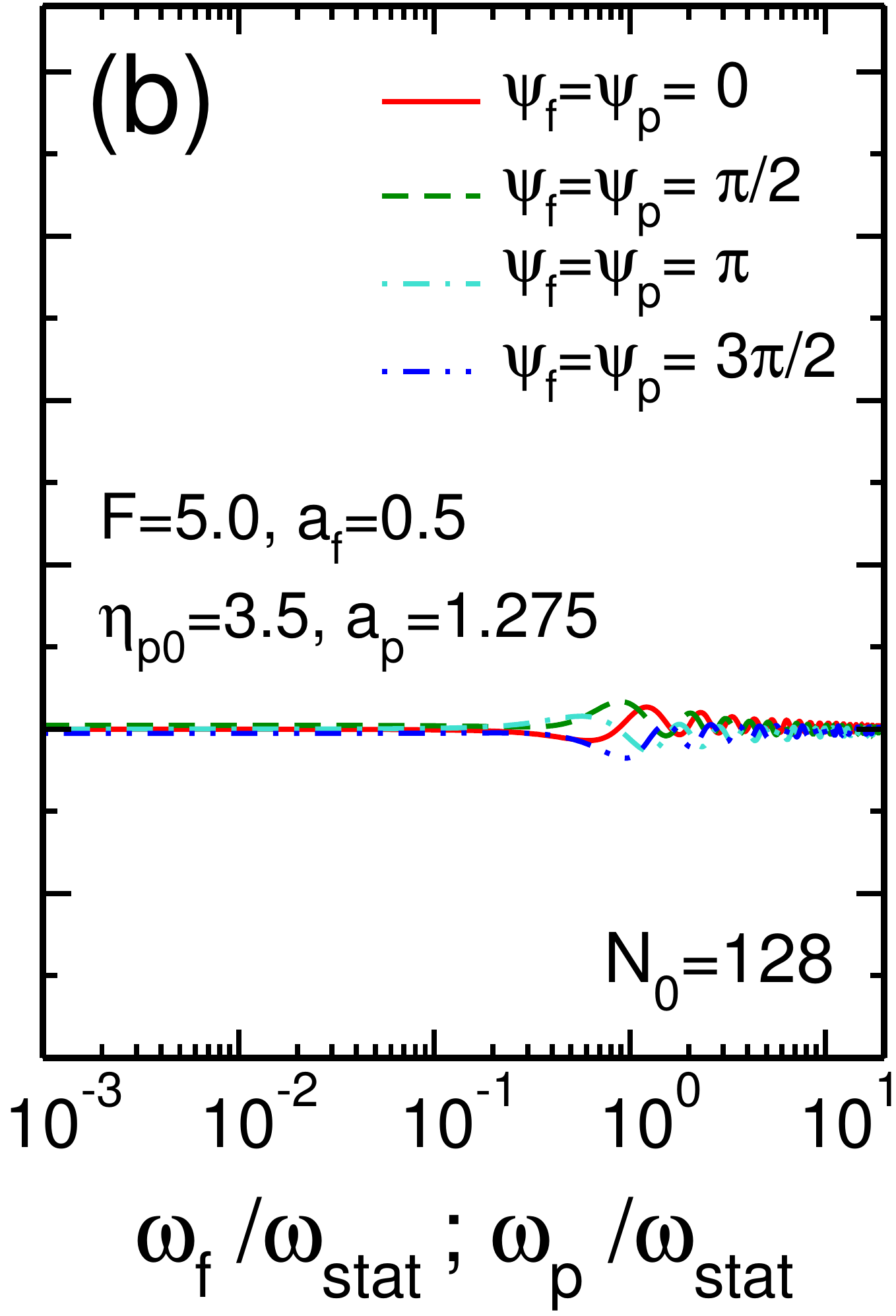}
    \end{center}\end{minipage}
\caption{(a) The normalized translocation time, $\tilde{\tau} / \tilde{\tau}_{\textrm{stat}}$, as 
a function of either normalized force or pore frequency, 
$\omega_{\textrm{p}} / \omega_{\textrm{stat}} = \omega_{\textrm{p}} / \omega_{\textrm{stat}}$, 
for various values of the mixed initial phases $\psi_f= \pi/2$ and $\psi_{\textrm{p}}= 3\pi/2$ (red solid line), $\psi_f= 3\pi/2$
and $\psi_{\textrm{p}}= \pi/2$ (green dashed line), $\psi_f= \pi/2$ and $\psi_{\textrm{p}}= \pi$ (turquoise dashed-dotted line), 
and $\psi_f= \pi$ and $\psi_{\textrm{p}}= \pi/2$ (blue dashed-dotted-dotted line).
Here, both the force and the pore oscillate periodically.
Panel (b) is the same as panel (a) but for different values of the mixed initial phases $\psi_f= \psi_{\textrm{p}}= 0$ (red solid line), 
$\psi_f= \psi_{\textrm{p}}= \pi/2$ (green dashed line), $\psi_f=\psi_{\textrm{p}}= \pi$ (turquoise dashed-dotted line), 
and $\psi_f= \psi_{\textrm{p}}= 3\pi/2$ (blue dashed-dotted-dotted line).
} 
\label{Translocation_time_pore_driven_flexible_flickering_fig}
\end{center}
\end{figure*}
%


\subsection{Translocation time and scaling of the translocation time} \label{translocation_time_felxible_flickering}

To see how sensitive the average translocation time is to the initial value of the force and the pore friction (as determined by
the corresponding phase factors at $t=0$), in Fig. \ref{Translocation_time_pore_driven_flexible_flickering_fig}(a) 
we plot the normalized translocation time $\tau / \tau_{\textrm{stat}}$ as 
a function of either the normalized pore or force frequencies, 
i.e. $\omega_{f} / \omega_{\textrm{stat}} = \omega_{\textrm{p}} / \omega_{\textrm{stat}}$, 
for various values of the mixed initial phases $\psi_f= \pi/2$ and $\psi_{\textrm{p}}= 3\pi/2$ (red solid line), $\psi_f= 3\pi/2$
and $\psi_{\textrm{p}}= \pi/2$ (green dashed line), $\psi_f= \pi/2$ and $\psi_{\textrm{p}}= \pi$ (turquoise dashed-dotted line), 
and $\psi_f= \pi$ and $\psi_{\textrm{p}}= \pi/2$ (blue dashed-dotted-dotted line).
The force and the pore friction are given by
Eqs. (\ref{oscillatory_force}) and (\ref{flickering_pore}), respectively, with the same parameters as in 
Fig. \ref{WT_trans_time_pore_driven_flexible_flickering_fig}.

To explain the influence of the oscillating quantities to translocation dynamics we should consider the different limits of the
problem. In the limit of low frequencies, the process of translocation occurs during the first half of the
oscillating force and/or oscillating pore friction period. 
For $\psi_f= \pi/2$ and $\psi_{\textrm{p}}= 3\pi/2$ (red solid line) during this first half of the
cycle, the value of the force decreases from its maximum value to its minimum, while the value of the pore friction increases 
from its minimum to its maximum. Therefore, the translocation time gradually increases for small frequencies and shows a maximum 
at $\tilde{\omega}_{f}/\tilde{\omega}_{\textrm{stat}} = 0.5$.
For frequencies higher than $\tilde{\omega}_{f}/\tilde{\omega}_{\textrm{stat}} = 0.5$, the polymer chain starts feeling the second half of
the cycle where the value of the force is increasing from its minimum value, while the pore friction is decreasing from its maximum value, 
and this leads to a first minimum at $\tilde{\omega}_{f}/\tilde{\omega}_{\textrm{stat}} = 1.0$. For higher forces or pore friction frequencies,
i.e., $\tilde{\omega}_{f}/\tilde{\omega}_{\textrm{stat}} > 1$, 
the polymer chain again experiences the next half of the cycle, i.e., $\tilde{T}_{f} < \tilde{t} < 3 \tilde{T}_{f} /2 $, where 
$\tilde{T}_{f} = 2 \pi / \tilde{\omega}_{f}$. Here again the value of the force is smaller than its maximum value, while the value of the 
pore friction is greater than its minimum and thus the translocation time increases. 
As the frequency increases further the translocation time oscillates between minima and maxima with a decreasing
amplitude upon approaching the limit of the high frequency, where the average of the rapidly oscillating force component sums to
zero within the translocation time.

Figure 8(b) corresponds to panel (a) but for different values of the mixed initial phases $\psi_f= \psi_{\textrm{p}}= 0$ (red solid line), 
$\psi_f= \psi_{\textrm{p}}= \pi/2$ (green dashed line), $\psi_f=\psi_{\textrm{p}}= \pi$ (turquoise dashed-dotted line), 
and $\psi_f= \psi_{\textrm{p}}= 3\pi/2$ (blue dashed-dotted-dotted line). 
Figure \ref{Translocation_time_pore_driven_flexible_flickering_fig}(a) clearly shows that for small values of the frequencies, 
i.e. $\omega_{f} , \omega_{\textrm{p}} < \omega_{\textrm{stat}}$, the translocation time is very sensitive to the selection of the initial phases.
In contrast, in panel (b) the translocation time is insensitive to the initial phases for all range of the frequencies. 
This happens because for all the curves in panel (b) the two effects of the driving force and pore friction now work against each other, 
and an almost complete cancellation occurs in the low and high frequency limits.

Following subsection \ref{translocation_time_felxible_static_subsection} the 
scaling form of the translocation time is written as
\begin{equation}
\tilde{\tau} 
= \tilde{\tau}_{\textrm{stat}}
\!-\! \frac{1}{\tilde{f}} \int_{0}^{\tilde{\tau}}\!\!\! \tilde{A}_f (\tilde{t}) d\tilde{t}
+  \frac{1}{\tilde{f}} \int_{0}^{N_0} \!\!\! \tilde{A}_{\textrm{p}} (\tilde{t}) dN.
\label{scaling_trans_time_SS}
\end{equation}
One can show that in the high pore friction frequency limit $\tilde{\omega}_{\textrm{p}} \gg 1/\tilde{\tau}$,
for very long chains, and in the low pore friction frequency limit $\tilde{\omega}_{\textrm{p}} \ll 1/\tilde{\tau}$, 
the total translocation time is given by
\begin{equation}
\hspace{-0.3cm}
\tilde{\tau} \!=\! \tilde{\tau}_{\textrm{stat}}
\!-\! \left\{
 \begin{array}{c l}
\hspace{-0.15cm} \int_{0}^{\tilde{\tau}}\!\!\! \tilde{A}_f (\tilde{t}) d\tilde{t} / \tilde{f} , \hspace{+2.35cm} \tilde{\omega}_{\textrm{p}} \gg 1/\tilde{\tau};\\
\\
\hspace{-0.15cm}  [ \int_{0}^{\tilde{\tau}}\!\!\! \tilde{A}_f (\tilde{t}) d\tilde{t} 
\!-\! \tilde{a}_{\textrm{p}} \sin (\psi_{\textrm{p}} )N_0 ] / \tilde{f} \!,\! \hspace{+0.1cm} \tilde{\omega}_{\textrm{p}} \! \ll \! 1/\tilde{\tau}.
\end{array}\right.
\label{scaling_trans_time_SS_high_and_low_frequency_flickering}
\end{equation}
As can be seen in Eq. (\ref{scaling_trans_time_SS_high_and_low_frequency_flickering}), at the low frequency limit the behavior of the pore 
is similar to a static pore with pore friction of $\tilde{\eta}_{\mathrm{p}} + \tilde{a}_{\mathrm{p}} \sin (\psi_{\mathrm{p}} )$.
More details for the other limits of the scaling form of the translocation time and the translocation exponent can be found in Ref. \cite{jalal2015}.


\section{Summary and Conclusions} \label{conclusion}

In this paper we have presented a brief review on recent theoretical progress on the dynamics of driven translocation
of polymers thorough a nanopore. In the past, there have been many attempts to explain the driven translocation process by simple
scaling arguments or linear response theories such as the Fokker-Planck equation. However, at least for moderate and high driving
the correct theory of driven translocation processes is based on the combination of non-equilibrium
tension propagation on the {\it cis} side of the translocating chain and an iso-flux assumption of the monomer density at
the pore. These ideas can be combined into an analytic IFTP theory, which is based on the dynamics of the translocation coordinate
of the chain and the time-dependent friction due to tension propagation. 

In this review we have shown how the IFTP theory can
be applied to a variety of driven translocation problems, including the pore-driven and end-pulled cases. It yields exact
analytic scaling forms in the appropriate limits, and reveals the role of the various frictional terms in translocation dynamics in terms
of correction terms to asymptotic scaling. Such correction terms that are significant for typical parameters used in most computer
simulations to date explain why many different values for the scaling exponents have been reported in the literature. For the
pore-driven case the correct asymptotic exponent is $\alpha = 1 + \nu$, while for the end-pulled case is becomes $\alpha = 2$.
Further, we have demonstrated that in cases where the time dependence of the {\it trans} side friction cannot be neglected,
the IFTP theory can still be retained by augmenting the total friction with the {\it trans} side contribution. Unfortunately, 
for pore-driven case of a semi-flexible chain
there is no analytic description available to date for such terms.

The main ingredient missing in the current version of the IFTP theory is the influence of hydrodynamic interactions.
Preliminary work has shown, however, this may only affect effective scaling exponents while the
asymptotic scaling forms should still be valid \cite{ikonen2013}. However, it would be interesting to study this issue
more thoroughly. It is also of interest to consider additional translocation
scenarios where the IFTP theory could be applied, such as the combination of pore driving and end-pulling \cite{Santtu_EPJE_2009},
double-sided pulling \cite{jalalDoubleSide_2018}, translocation of hairpin loops, just to mention a few. We plan to work on these problems in the future.

\begin{acknowledgments}
JS acknowledges Timo Ikonen for enlightening discussions.
This work was supported in part by the Academy of Finland through its Centers of Excellence program under project Nos. 251748, 284621
and 312298. The numerical calculations were performed 
using computer resources from the Aalto University School of Science ``Science-IT'' project, and from CSC - Center for 
Scientific Computing Ltd. The MD simulations were performed with the LAMMPS simulation software.
\end{acknowledgments}


\end{document}